\documentclass{natureprintstylev4}
\usepackage{astjnlabbrev-nature}
\usepackage{hyperref}
\usepackage[switch]{lineno}

\usepackage{amssymb}
\usepackage{mathptmx}
\usepackage{amsmath}	
\usepackage{gensymb}
\usepackage{color}
\usepackage{enumitem}
\usepackage{textcomp}
\usepackage{multirow}
\usepackage{multibib}

\usepackage{epsfig}
\usepackage{color}
\usepackage{graphicx}
\usepackage{longtable}
\usepackage{hyperref}
\usepackage{soul}
\usepackage{ulem}

\usepackage[switch]{lineno}

\usepackage[labelsep=endash]{caption}

\newcommand{\kms}{km\,s$^{-1}$}

\newcommand{\thestar}{Lan 11}

\newcommand{\sdB}{_\textrm{sd}} 
\newcommand{\comp}{_\textrm{comp}} 

\newcommand{\Gaia}{{\it Gaia}}

\newcommand\arcsec{\mbox{$^{\prime\prime}$}}
\newcommand{\lijiao}[1]{{#1}}
\newcommand{\zcj}[1]{{#1}}

\newcommand{\citep}{\cite}
\newcommand{\citet}{\cite}

\title{A born ultramassive white dwarf-hot subdwarf super-Chandrasekhar candidate}

\author{Changqing Luo$^{1,8}$, Jiao Li$^{1,8}$, Chuanjie Zheng$^{1,2,8}$, Dongdong Liu$^{3}$, Zhenwei Li$^{3}$, Yangping Luo$^{4}$, P{\'e}ter N{\'e}meth$^{5,6}$, Bo Zhang$^{1}$, Jianping Xiong$^{1}$, Bo Wang$^{3}$, Song Wang$^{1}$, Yu Bai$^{1}$, Qingzheng Li$^{3,2}$, Pei Wang$^{1,7}$, Zhanwen Han$^{3}$, Jifeng Liu$^{1,2}$, Yang Huang$^{2,1,9}$, Xuefei Chen$^{3,9}$, Chao Liu$^{1,2,9}$}

\begin{document}
\setstcolor{red}

\maketitle

\begin{affiliations}
\item National Astronomical Observatories, Chinese Academy of Sciences, Beijing 100101, People's Republic of China;\\
\item School of Astronomy and Space Science, University of Chinese Academy of Sciences, Beijing 100049,  People's Republic of China;\\
\item Yunnan Observatories, Chinese Academy of Sciences, Kunming 650011, People's Republic of China;\\
\item Department of Astronomy, China West Normal University, Nanchong, 637002, People's Republic of China; \\
\item Astronomical Institute of the Czech Academy of Sciences, CZ-251 65, Ond{\v{r}}ejov, Czech Republic;\\
\item Astroserver.org, F\H{o} t\'{e}r 1, 8533 Malomsok, Hungary;\\
\item{Institute for Frontiers in Astronomy and Astrophysics, Beijing Normal University, Beijing 102206, China}\\
\item These authors contributed equally to this work.\\
\item Corresponding authors: huangyang@bao.ac.cn; cxf@ynao.ac.cn; liuchao@nao.cas.cn
\end{affiliations}
\\


\begin{abstract}
 {Although supernovae is a well-known endpoint of an accreting white dwarf, alternative theoretical possibilities has been discussing broadly, such as }the accretion-induced collapse (AIC) event 
 {as the} endpoint of oxygen-neon (ONe) white dwarfs\citep{Miyaji80}$^{,}$\citep{Nomoto84}, either accreting up to or merging to excess the Chandrasekhar limit (the maximum mass of a stable white dwarf).
AIC is an important channel to form neutron stars, especially for those unusual systems,  {which are} hardly produced by core-collapse supernovae\citep{Tauris13}$^{,}$\citep{LL17}.
However, the observational evidences for this theoretical predicted event and its progenitor are all very limited\citep{WL20}. 
 {In all of the known progenitors, white dwarfs increase in mass by accretion.}
Here, we report the discovery of an intriguing binary system Lan\,11, consisted of a stripped core-helium-burning hot subdwarf and  {an unseen compact object of $1.08$ to $1.35$ $M_{\odot}$. 
Our binary population synthesis calculations, along with the absence of detection from the deep radio observations of the Five-hundred-meter Aperture Spherical Radio Telescope, strongly suggest that the latter is an ONe white dwarf.}
The total mass of this binary is { $1.67$ to $1.92$ $M_{\odot}$},  significantly excessing the Chandrasekhar limit.
The reproduction of its evolutionary history indicates that the unique system has undergone two phases of common envelope ejections, implying a born nature of this massive ONe white dwarf rather than an accretion growth from its companion.
These results, together with short orbital period of this binary (3.65 hours), suggest that this system will merge in  {$500-540$\,Myr, largely} triggering an AIC event\citep{Liu20},  {although the possibility of type Ia supernova cannot be fully ruled out\citep{Marquardt2015}. This finding greatly  provides valuable constraints on our understanding of stellar endpoints, whatever leading to an AIC or a supernova.}
\end{abstract}

\section*{Main}
 {As a special binary system, unlike the usually discussed binary system with a companion of carbon-oxygen (CO) white dwarf (WD), the final fate of a binary star with a massive oxygen-neon (ONe) white dwarf approaching the Chandrasekhar limit could be either the accretion-induced collapse (AIC)\citep{Miyaji80}$^{,}$\citep{Nomoto84} or supernovae type Ia\citep{Marquardt2015}.}
Depending on the formation channels, AIC has a wide range of delay time, from several tens of Myr to $> 10$\,Gyr, after the birth of its progenitor, and thus can naturally explain the very young radio pulsars found in globular clusters that can not be produced by the core-collapse supernovae given the massive nature of its progenitor\citep{Boyles11}.
The AIC scenario can also be responsible for explaining a number of unusual neutron stars (NSs), e.g.  recycled pulsars with low space velocities due to their low kick velocities\citep{BG90}, intermediate-mass binary pulsars with short orbital periods\citep{Liu18}, the strong magnetic field and slow spin NSs with ultra-light companions ($\leq\,0.1\,M_{\odot}$) in close orbits\citep{Tauris13}$^{,}$\citep{Liu23}, and can even be a promising progenitor of fast radio burst\citep{Margalit19} (FRB).
The AIC however is still a theoretical prediction without conclusive evidences in observations, although massive efforts are spent to search for both the explosion events and its progenitors\citep{WL20, Ruiter2019MNRAS.484..698R}.

\begin{table*}[ht!]
\centering          
\begin{tabular}{c c c c c c} 
\hline\hline       
Parameter & \multicolumn{5}{c}{Value}\\
\hline
     R.A. (J2000) & \multicolumn{5}{c}{06:00:30.98}  \\
     Decl. (J2000) & \multicolumn{5}{c}{+29:08:55.06} \\
     $G$ (mag) & \multicolumn{5}{c}{$13.079\pm0.001$} \\
     $\varpi$ (mas) & \multicolumn{5}{c}{$1.146 \pm0.029$} \\
     $d$ (pc)  & \multicolumn{5}{c}{$873\pm22$} \\ 
     $\mu_{\alpha}$ (mas\ yr$^{-1}$) & \multicolumn{5}{c}{$-1.96\pm0.03$} \\
     $\mu_{\delta}$ (mas\ yr$^{-1}$) &  \multicolumn{5}{c}{$-2.91\pm0.02$} \\
     $P$ (min)     & \multicolumn{5}{c}{219.08880(40)} \\
     RUWE &   \multicolumn{5}{c}{0.9920} \\
\hline
\hline
& \multicolumn{2}{c}{Spectroscopic} & \multicolumn{2}{c}{SED} & Light curve\\
 & phase-picked & highest SNR & phase-picked & highest SNR & Adopted interval \\
\hline
    $T_{\rm eff}^{\rm sd}$ (K) & $35840\pm130$ & $35850\pm140$ &$35831\pm128$ &$35839\pm140$ & [35710, 35992] \\
	$\log g_{\rm sd}$ [cm s$^{-2}$] & $5.306\pm0.020$ & $5.338\pm0.020$& $5.302\pm0.020$& $5.334\pm0.020$& [5.292, 5.365] \\
    $\log n{\rm He}/n{\rm H}$ & $-0.867\pm0.033$ & $-0.937\pm0.028$& \multicolumn{2}{c}{--} & --\\
    $E(B-V)$ (mag)&\multicolumn{2}{c}{--} & 0.16056(89) & 0.16069(89) & --\\
    $R_{\rm sd}$ ($R_\odot$) & \multicolumn{2}{c}{--} &\multicolumn{2}{c}{$0.275\pm0.007$} & [0.265, 0.282] \\
\hline
	$T_0$ (BJD-2459477)   & \multicolumn{2}{c}{0.82316(46)} & \multicolumn{2}{c}{--} & [0.82161, 0.82195] \\
	$K_{\rm sd}$ (\kms) &\multicolumn{2}{c}{$249.8\pm2.1$} & \multicolumn{2}{c}{--} &  [247.6, 252.2] \\
    $RV_{\rm \gamma}$ (\kms) &\multicolumn{2}{c}{$-4.8\pm1.5$} & \multicolumn{2}{c}{--} & --  \\
	$\sqrt{e}\cos\omega$ &\multicolumn{2}{c}{$-0.148\pm0.035$} & \multicolumn{2}{c}{--} &  [$-0.070$, 0.015] \\
	$\sqrt{e}\sin\omega$ &\multicolumn{2}{c}{$0.039\pm0.052$} & \multicolumn{2}{c}{--} &  [$-0.023$, 0.038]\\
\hline
    $i$ (\degree)  & \multicolumn{2}{c}{--}& \multicolumn{2}{c}{--}  & [45.4, 55.5]\\
    $M_{\rm sd}$ ($M_\odot$) & \multicolumn{2}{c}{--} &\multicolumn{2}{c}{--} & [0.524, 0.651]\\
	$M\comp$ ($M_\odot$) & \multicolumn{2}{c}{--} &\multicolumn{2}{c}{--}& [1.0751, 1.353] \\
    $M_{\rm all}$ ($M_\odot$) & \multicolumn{2}{c}{--} &\multicolumn{2}{c}{--}& [1.673, 1.922] \\
    $q$           & \multicolumn{2}{c}{--} & \multicolumn{2}{c}{--} & [1.73, 2.48]  \\
    $L_{\rm sd}$ ($L_\odot$) & \multicolumn{2}{c}{--} &\multicolumn{2}{c}{--}& [104.5, 118.4] \\
    $a$ ($R\odot$) & \multicolumn{2}{c}{--}& \multicolumn{2}{c}{--} & [1.423, 1.491] \\
	$R^{\rm Roche-lobe}_{\rm sd}$ ($R_{\odot}$) & \multicolumn{2}{c}{--}& \multicolumn{2}{c}{--} &  [0.445, 0.482] \\
	$R^{\rm Roche-lobe}\comp$ ($R_{\odot}$) & \multicolumn{2}{c}{--}& \multicolumn{2}{c}{--} &  [0.612, 0.681] \\
   \hline
\end{tabular}
   \caption{\textbf{Astrometric, stellar, and orbital parameters for \thestar{}}.  Astrometric parameters are taken from {\it Gaia} DR3\cite{Gaiadr3}. The reported parallax has been corrected for a zero-point of $-0.033$\,mas\cite{Lindegren2021AA}.
   The stellar parameters are mainly determined from the spectral template matching and the SED fitting.
   The orbital parameters are constrained from the radial velocity curve fitting and TESS light curve modeling by using {\tt ellc} and {\tt PHOEBE} independently.
   Quantities shown are the right ascension RA, declination DEC, \Gaia{} $G$-band magnitude, parallax $\varpi$, distance $d$, proper motions in the {directions of }right ascension $\mu_\alpha$, and declination $\mu_\delta$, the orbital period $P$, the Renormalised Unit Weight Error RUWE, the hot subwarf effective temperature $T_{\rm eff}^{\rm sd}$, surface gravity $\log~g_{\rm sd}$, helium abundance $\log n{\rm He}/n{\rm H}$, extinction $E(B-V)$, radius $R_{\rm sd}$, the superior conjunction time $T_0$, the radial velocity semiamplitude of the hot subdwarf $K\sdB$, the systemic velocity of the binary $RV_\gamma$, the eccentricity $e$, the longitude of periastron $\omega$, the orbital inclination $i$, the hot subdwarf star mass $M_{\rm sd}$, the companion star mass $M\comp$, the total mass of the binary $M_{\rm all}$, the mass ratio $q$, luminosity $L\sdB$, the orbital separation $a$, the effective Roch-lobe radii of the hot subdwarf star $R^{\rm Roch-lobe}\sdB$ and the companion star $R^{\rm Roch-lobe}\comp$. 
   }
   \label{tab:binary2}
\end{table*}

In this study, we report the discovery of a hot subdwarf–WD binary LAN\,11 as a promising AIC progenitor. 
This system is identified as a nearby ($873$\,pc to the Sun, which is estimated from the {\it Gaia} parallax measurements\citep{Gaiadr3}$^{,}$\citep{Lindegren2021AA}), bright ({ visual magnitude} $V = 13.19$) hot subdwarf by the LAMOST low-resolution spectra\citep{Lei18}, with large variations in radial velocities revealed by the multi-epoch LAMOST medium-resolution spectroscopic observations.
The atmospheric and orbital parameters (listed in Table\,1), especially the mass of companion, of this system are then well determined by the high-quality follow-up time-series spectroscopy, from Palomar 200-inch telescope, together with the precise light curves obtained by the Transiting Exoplanet Survey Satellite (TESS).
The result shows this is the most massive one among the three detected super-Chandrasekhar hot subdwarf–WD binary systems\citep{Maxted00}$^{,}$\citep{Pelisoli21}.
 {It is the first ever found non-accretion binary system with massive WD companion, contrasting with the other two massive white dwarfs found in novae V603 Aql (ref.\citep{Arenas00}) and RN U Sco (ref.\citep{Thoroughgood01}), which are accreting material from their companions. The heavy mass greater than 1.1\,$M_{\odot}$ of the WD indicates that this system will largely end up as an AIC rather than a type Ia supernova (SNe Ia), although the possibility of the latter cannot be totally excluded\citep{Marquardt2015}.}

\section*{Results}

\begin{figure*}
    \centering
    \includegraphics[scale=0.3, angle=0]{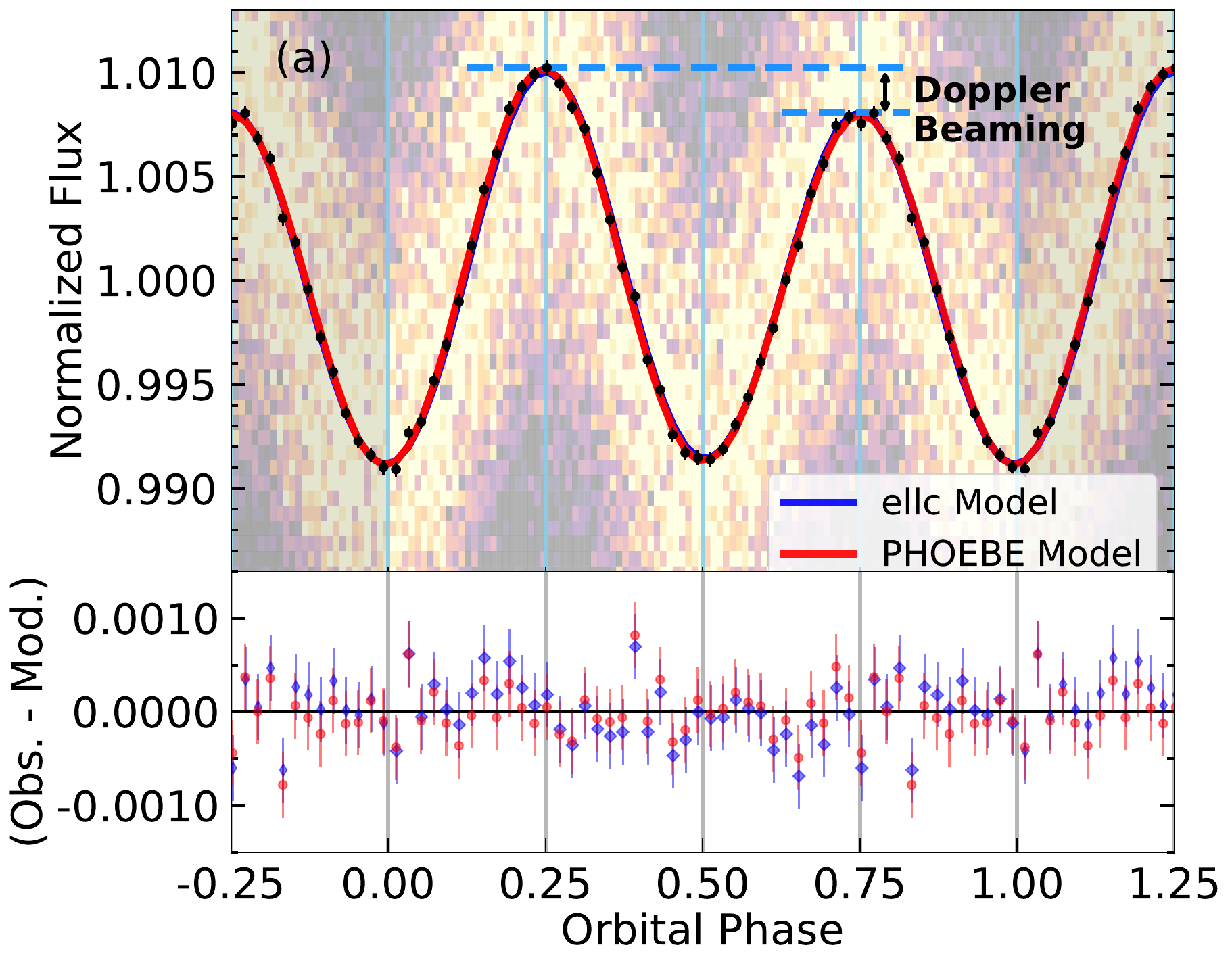}
    \includegraphics[scale=0.3, angle=0]{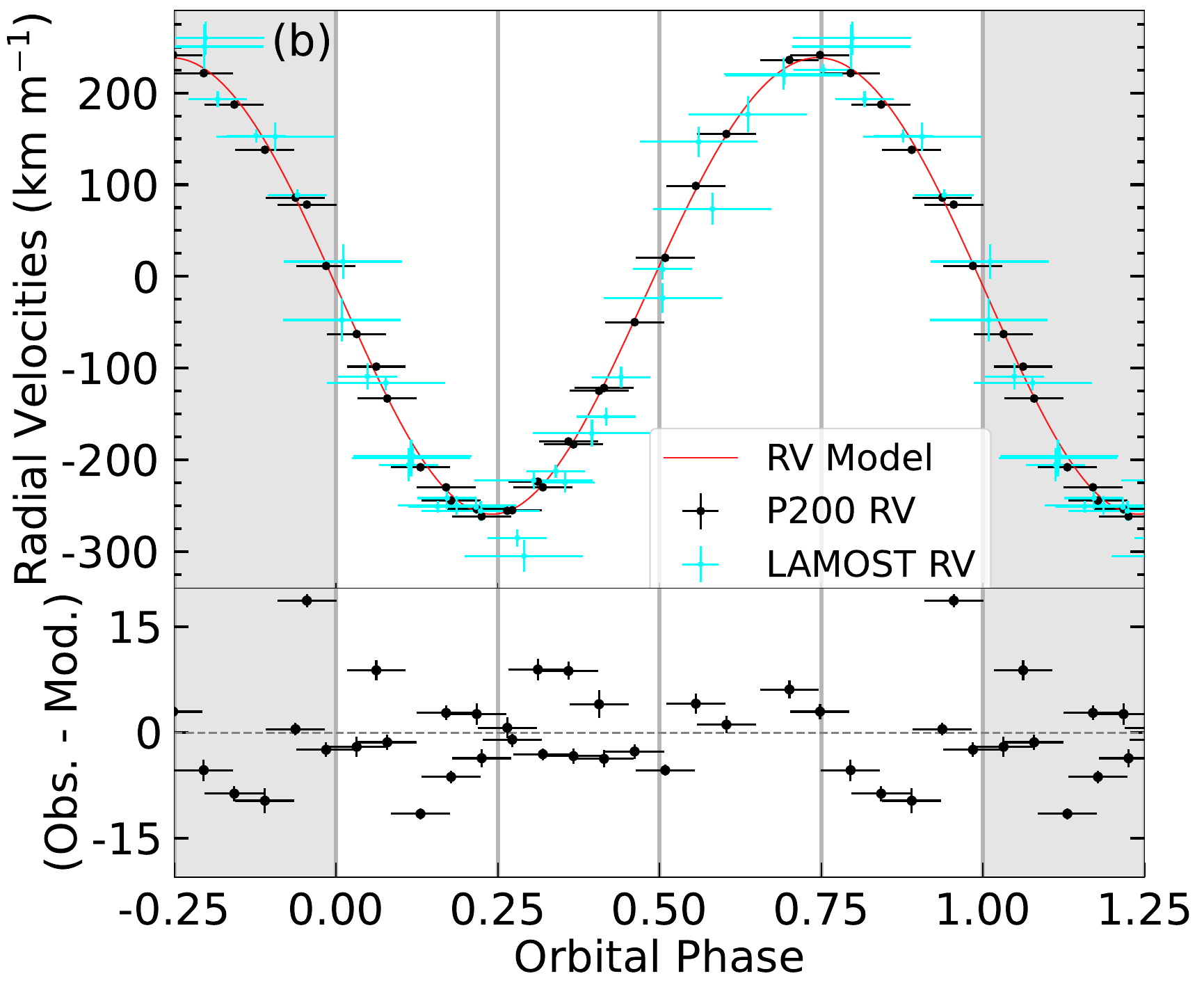}   
    \caption{\textbf{Phased data for Lan\,11}. \textbf{(a):} Phase folded light curve and its fitting residuals. The original TESS data points are shown in the background density plot. Black dots represent the binned data by each one-fifty phase for the light curve fit. The error bars show the $1\sigma$ uncertainty for each binned data.
    The different height between adjacent maxima is mainly caused by the Doppler boosting.
    The red and blue solid lines represent, respectively, the best-fit models found by {\tt ellc} (blue) and {\tt PHOEBE} (red)  {with priors on atmospheric parameters from the highest SNR spectrum}.
    The lower part of this panel shows the model residuals.
    \textbf{(b):} Phase folded radial velocity curve and its fitting residuals.
    The individual radial velocity measurements are shown as black dot (from P200) and cyan dot (from LAMOST), respectively, as well as their $1\sigma$ uncertainties {and exposure duration}.
    The red line represents the best-fit model for the radial velocity curve obtained from P200, with the fitting residuals shown in the lower part.}
    \label{fig:lc-rv}
\end{figure*}

\textbf{Light curve.} 
Lan\,11 was observed by TESS in Sectors 43, 44, and 45 with 2 minutes cadences.
The whole data, including 50,447 data points, covers a baseline of around 76.5 days.
A significant peak at 109.54440(20) min is seen in the periodogram by performing Lomb-Scargle analysis to the TESS light curve (see Supplementary Figure~\ref{fig:lc_period}).
The phase-folded light curve, assuming a period of twice of the peak value found in the periodogram, is shown in Figure\,~\ref{fig:lc-rv}(a).
Clearly, a small but significant effect of Doppler beaming is detected between adjacent maxima, with a height difference {around 0.2\%}, suggesting an orbital period of 219.08880(40)\,min  {(corresponding to the second harmonic peak in Supplementary Figure~\ref{fig:lc_period})}.
The light curve also shows obvious ellipsoidal variations of a semi-amplitude of 1\% level, implying significant tidal force from a compact companion.  

\textbf{Radial velocity curve and stellar parameters}
Between October 2017 and February 2021, 34 exposures medium-resolution spectra with $R \sim 7500$ are collected for Lan\,11 by LAMOST (see Supplementary Table\,1).
The radial velocities are measured from the H${\rm \alpha}$ lines by using the LAMOST spectra from red arm (see Methods).
The phase-folded radial velocity curve, by adopting the orbital period obtained above, is presented in Figure\,~\ref{fig:lc-rv}(b), showing a significant sinusoid-like motions with semi-amplitude larger than 200\,km\,s$^{-1}$. 
Due to the shallow limiting magnitude of LAMOST medium-resolution observations, the quality, in sense of both time resolution (longer than 10 minutes) and measurement errors (typically 10\,km\,s$^{-1}$), of the radial velocity curve suffers from large uncertainties.
Time-series spectra are therefore collected for Lan\,11 by using the Double-Beam Spectrograph (DBSP) equipped in the Palomar 200-inch telescope during February 19, 2022 (see Supplementary  {Table\,2}).
In total, 28 spectra with a typical spectral signal-to-noise ratio (SNR) greater than 80 are obtained, covering the full orbital phase of Lan\,11.
The radial velocities are then precisely determined by performing cross-correlation of each collected spectrum to the model template (see Methods for details).
The phase-folded radial velocity curve from DBSP, with a typical measurement error of 1.3\,km\,s$^{-1}$, is also shown in Figure\,~\ref{fig:lc-rv}(b).

The canonical single-line binary radial orbit model is adopted to fit this radial velocity curve, by fixing the period to the solution of light curve (see above).
The best-fit model, shown in Figure\,~\ref{fig:lc-rv}(b), reveals a large radial semi-amplitude of
{$K_{\rm sd} = 249.8\pm2.1$}\,km\,s$^{-1}$ of this hot subdwarf. 
The mass function of the unseen companion is then determined by the light curve and radial orbital solutions:
\begin{equation}
\begin{split}
    f\comp &= \frac{M^3_{\rm comp}\sin^3 i}{(M_{\rm sd} +M_{\rm comp})^2}\\
           &= \frac{PK^3_{\rm sd}}{2\pi G}=0.2456_{-0.0061}^{+0.0063}.
\end{split}\label{eq:fbinary}
\end{equation}
The systemic velocity from the radial orbital solution, together with the positions, proper motions and distances measured from {\it Gaia}\citep{Gaiadr3}, shows that this system is rotating the Galactic center with a speed of  {$237.80^{+0.18}_{-0.20}$\,km\,s$^{-1}$}, implying its nature of the thin disk population (Methods).

\begin{figure*}
    \centering
    \includegraphics[scale=0.30, angle=0]{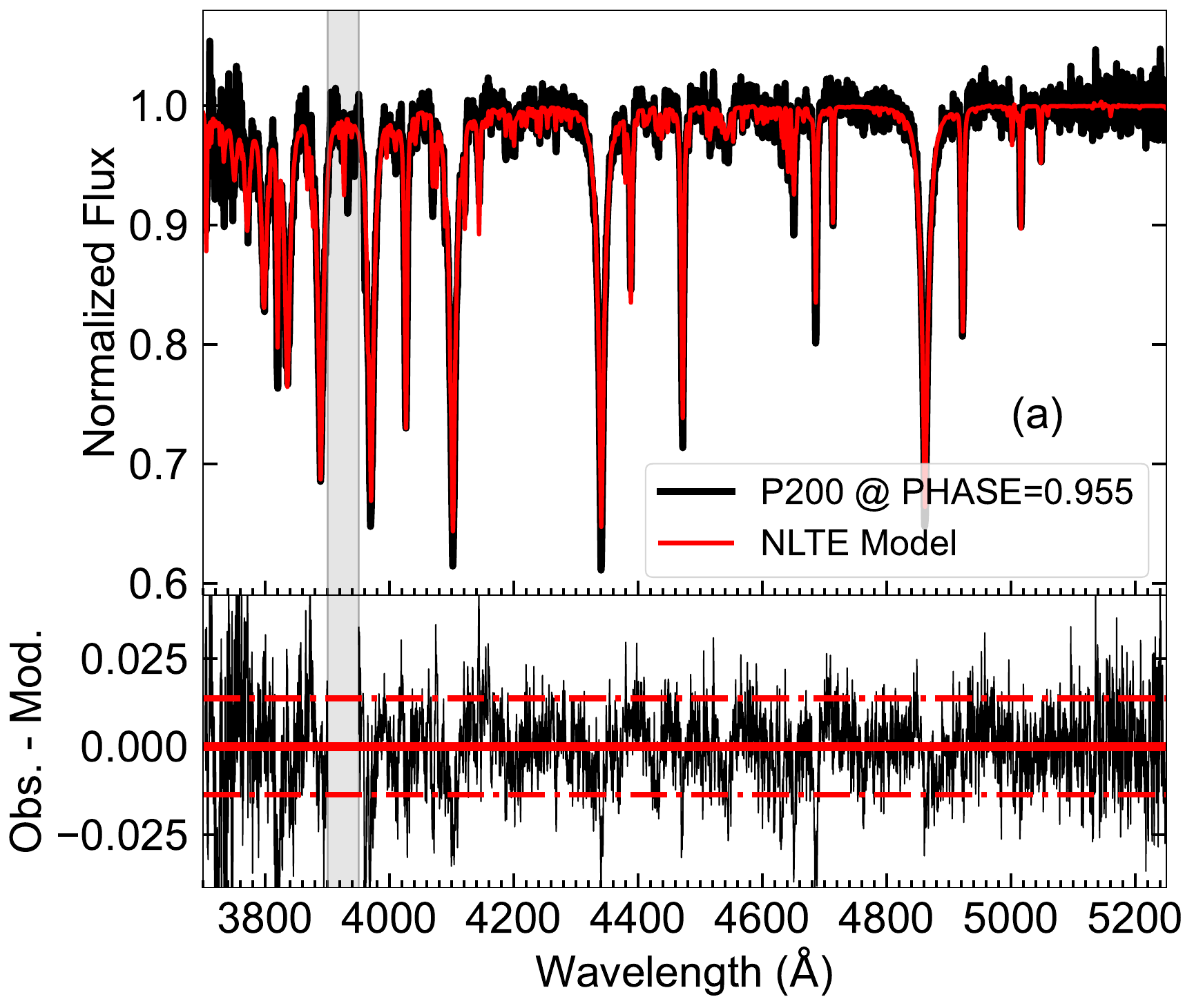}
    \includegraphics[scale=0.30, angle=0]{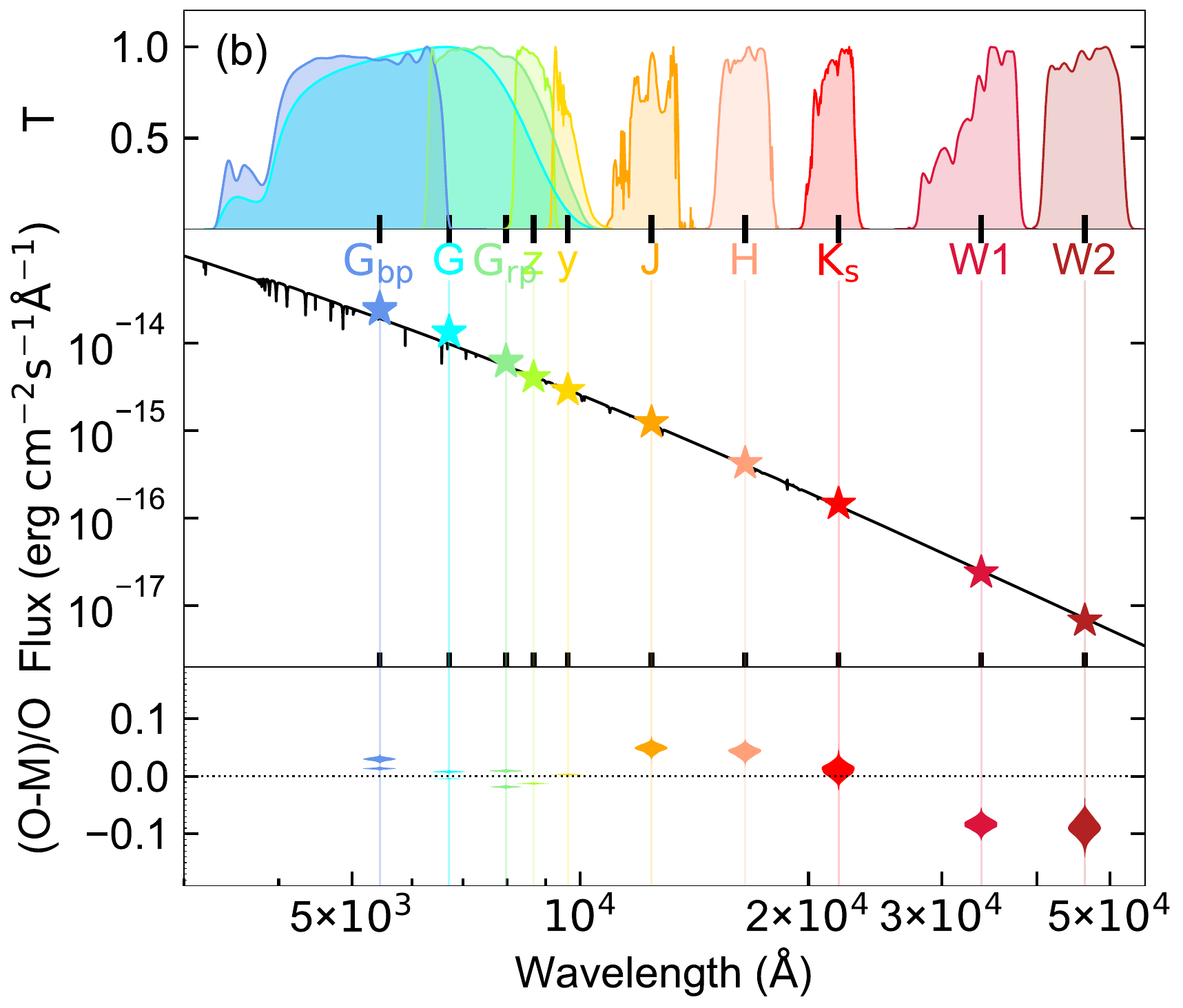}
    \caption{\textbf{The estimates of the stellar parameters of hot subdwarf.} \textbf{(a):} The normalized  {highest SNR DBSP} spectrum of Lan\,11 is shown as black line, together with the best-fit synthetic spectrum ($T_{\rm eff} = 35,850$\,K, log\,$g = 5.338$ and $\log n{\rm He}/n{\rm H} = -0.937$) shown in red.
    The fitting residuals are shown in the lower part of the panel. The dotted line represents the standard deviation of the residuals.
     {The shadow vertical band represents the Ca K line region ($3850-3885$\AA) that is masked (avoid the influence of interstellar medium absorption) during spectral fits.}
    \textbf{(b):} The broadband spectral energy distribution of Lan\,11.
    Stars represent the observed flux from the {\it Gaia} EDR3 $G_{\rm bp}$-, $G$- and $G_{\rm rp}$-bands, PanSTARRS $z$- and $y$-bands, 2MASS $J$-, $H$- and $K_{\rm s}$-bands, and ALLWISE $W_1$-$W_2$ bands. 
    The transmission curves for these braodband filters are displayed in the top panel and normalised at the maximum.
    The black curve is the best-fit `tmap' model spectrum for the hot subdwarf.
    The relative fitting residual, defined as (O$-$M)/O, is shown in the lower panel. {Vertical range of these violin symbols represent the photometric uncertainties.}}
    \label{fig:spec-sed}
\end{figure*}

By performing spectral fits to  {either the single-epoch DBSP spectrum with the highest signal-to-noise ratio (SNR) or four phase-picked high-SNR DBSP spectra} with the {\sc XTgrid} tool\citep{Nemeth19} (Methods), the visible component is found to be an sdOB with  {$T_{\rm eff}^{\rm sd}$ around 35,800\,K (with a typical uncertainty of 140\,K),  $\log g_{\rm sd}$ ranging from 5.29 to 5.36 and a helium-to-hydrogen ratio log\,$n{\rm He}/n{\rm H}$ between $-0.83$ and $-0.97$ (see Table\,1).}
As shown in Figure~\ref{fig:spec-sed}(a) and { Supplementary Figure~\ref{fig:dbsp_lines}}, the best-fit model spectrum agrees very well with the observed ones from DBSP.
The obtained parameters are  {also} consistent with previous estimates from LAMOST low-resolution spectra\citep{Lei18}. 

By combing with the distance estimate from {\it Gaia} parallax, we determined the stellar radius of the hot subdwarf from its spectral energy distribution (SED) composed of the multi-band photometry. The values of $T_{\rm eff}^{\rm sd}$ and log\,$g_{\rm sd}$, as well as uncertainties,  {yielded from above spectral fits} are taken as the priors during the SED fitting (see Methods).
 Figure~\ref{fig:spec-sed}(b) clearly shows that the contribution of a single hot subdwarf with $T_{\rm eff}^{\rm sd}$ and log\,$g_{\rm sd}$ listed in Table\,1 is in excellent agreement with the observed SED, implying negligible contributions from the unseen companion.
The resulting reddening of $E (B-V) = \zcj{0.161} \pm 0.001$ is very close to the value of $0.168$ reported by the 3D dust map\citep{Green19}.
We finally obtained the radius of  {$R_{\rm sd} = 0.275 \pm 0.007\,R_{\odot}$} for the hot subdwarf.

\begin{figure*}
    \centering
    \includegraphics[scale=0.25, angle=0]{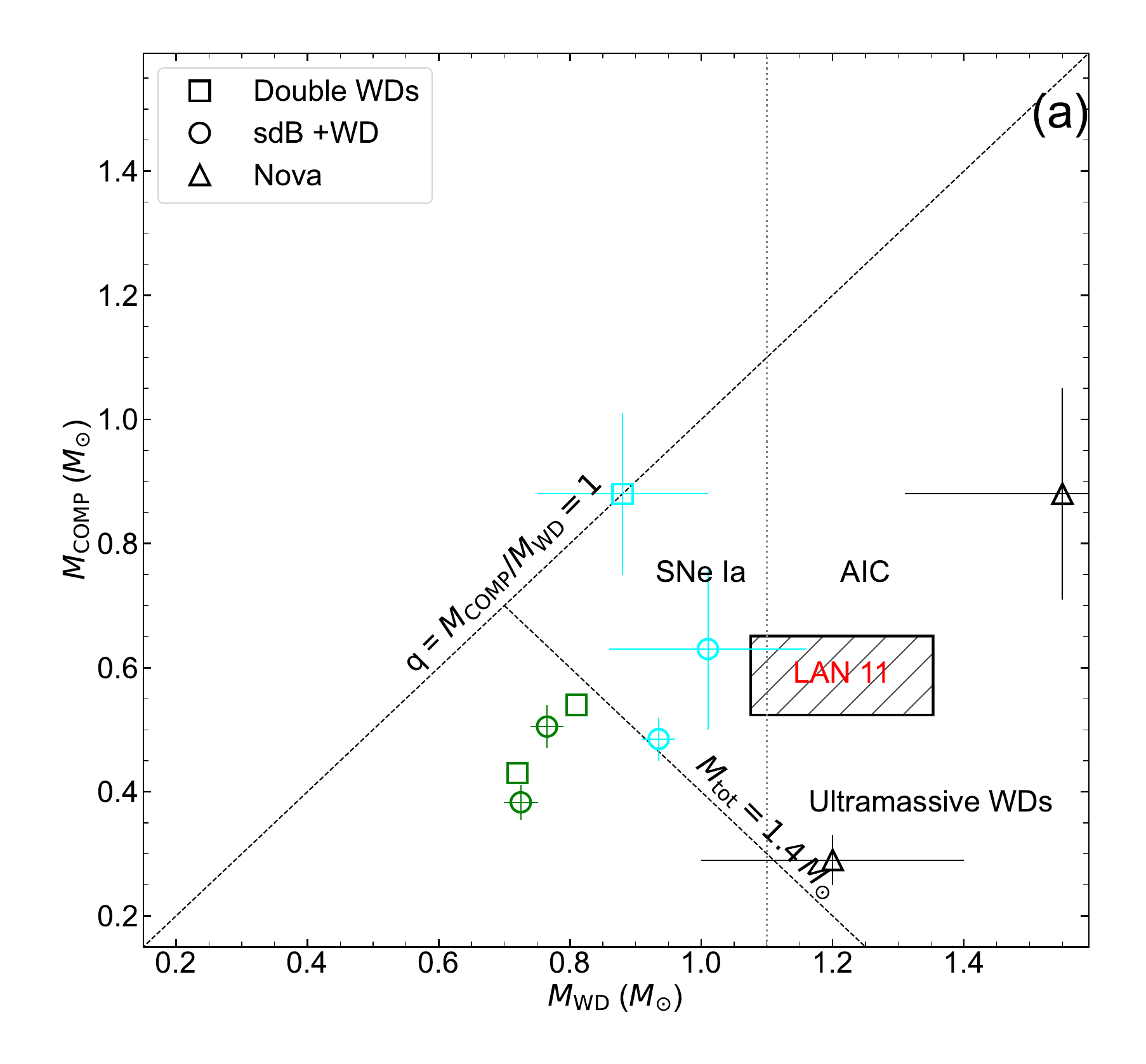}
    \includegraphics[scale=0.25, angle=0]{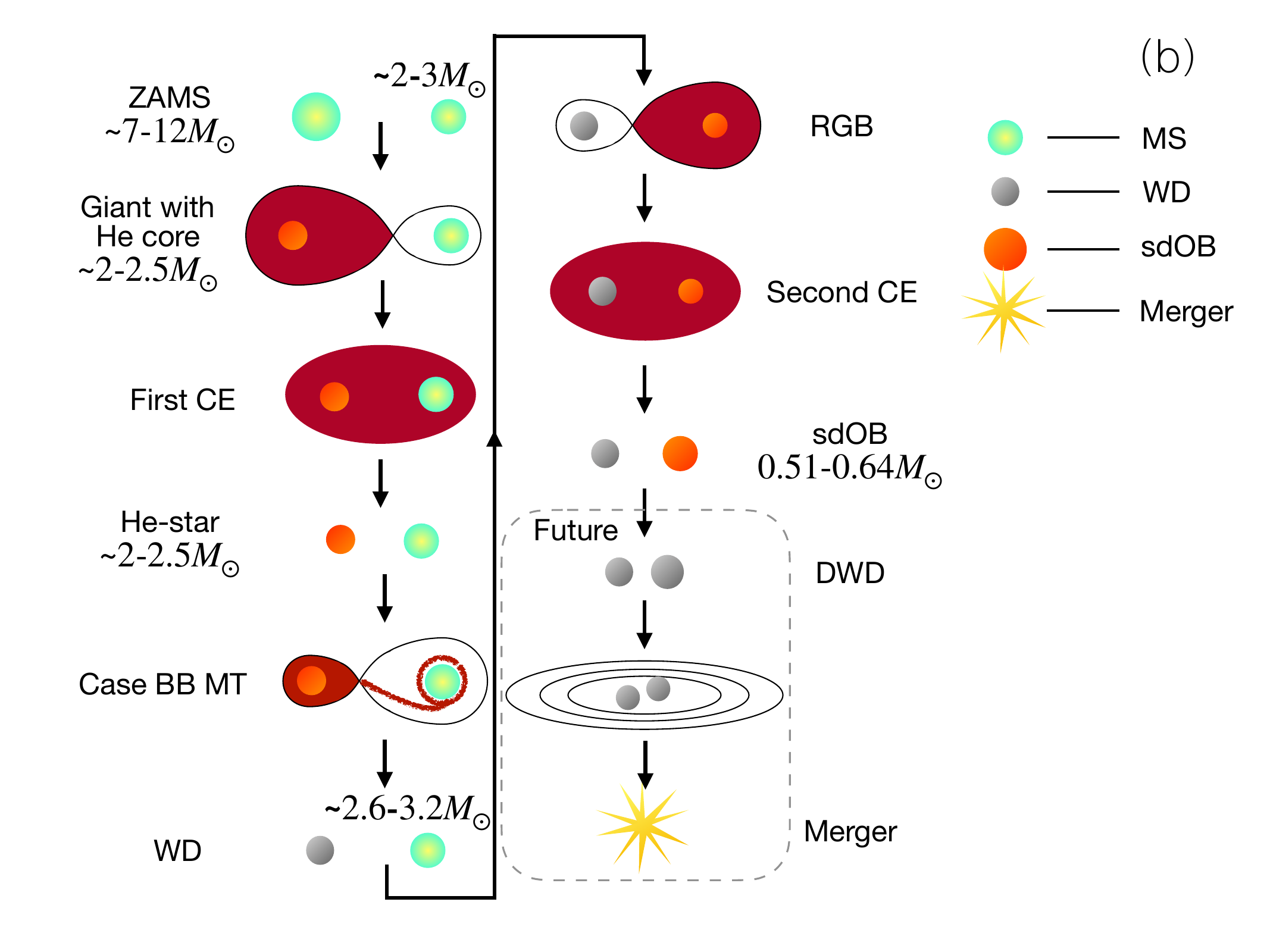}
    \caption{\textbf{The mass-mass diagram and the carton of the evolutionary history of Lan 11.} \textbf{(a):} Mass-mass diagram of the binary systems that are potential candidates of type Ia supernovae (SNe Ia) and accretion-induced collapse (AIC). Symbols of square, circle and triangle represent the double white dwarfs (WDs), the hot subdwarf (sdB)--WD and the classical or recurrent nova consisted by a WD plus a non-degenerate companion. The green and cyan symbols mark the known sub-Chandrasekhar and super-Chandrasekhar candidates of SNe Ia. The dotted line indicates the definition of ultramassive WD with mass greater than 1.1\,$M_{\odot}$.
    Lan\,11 is the only sdB-WD system that could end up with an AIC event. 
    \textbf{(b):} { The past and future of Lan\, 11. The binary may experience two CE ejection phases and a case BB mass transfer phase to produce Lan\,11. }
   }
    \label{fig:fate}
\end{figure*}

\section*{Discussion}
\textbf{The nature of the unseen companion.}
The ultra-high precision light curve from TESS has enabled us to infer the mass of the unseen companion star, given its modulations on the light curve from the tidal force.
We then modeled the light curve of Lan\,11 by two independent advanced software packages: {\tt ellc}\citep{Maxted16} and {\tt PHOEBE}\citep{PZ05} (Methods).
The detail setups, including priors of input parameters, are described in Methods.
The well-constrained parameters, including the inclination angle ($i$), the eccentricity ($e$), and the longitude of periastron ($\omega$), are consistent with each other from the light curve solutions of both {\tt ellc} and {\tt PHOEBE}.
By combing the solutions from both packages, we obtain an inclination angle of  {$45.4-55.5$} degree and a companion mass of  {$1.08-1.35$}\,$M_{\odot}$.
As suggested by our binary population synthesis (BPS) simulation (Methods),  {this unseen object}  {has more than $85$\% probability to be} 
{a WD with an ONe core, according to its massive nature (ref.\citep{Lauffer18}).}
The  {probability} that the massive companion is a neutron star (NS)  {is only around $5$\%,
meaning that this channel} can not be  {fully} ruled out. { The probability to be a NS is much lower than that of an ONe WD. This is due to that}
 {the high-mass (largely greater than 12\,$M_{\odot}$, ref.\citep{Toonen2018A&A...619A..53T}) of neutron star's progenitor is not preferred by both the initial mass function and the extremely mass-ratio of the primordial binary ($q \ge 4$; see Figure\,~\ref{fig:fate}(b)).
Moreover, the companion nature of a WD (rather than a NS) is supported by the non-detection from the deep radio observations (Methods) of the Five-hundred-meter Aperture Spherical radio Telescope (FAST).}


\textbf{The past and future of Lan 11}
This system, with a total mass of  {$1.67-1.92$}\,$M_{\odot}$, is the most massive super-Chandrasekhar hot subdwarf–WD ( {suggested by our BPS simulation)} binary system ever found (see Figure\,~\ref{fig:fate}(a)), which is very interesting and unique system compared to other super/sub-Chandrasekhar systems\citep{Maxted00}$^{,}$\citep{Arenas00}$^{,}$\citep{Thoroughgood01}$^{,}$\citep{Geier13}$^{,}$\citep{SG15}$^{,}$\citep{Pelisoli21}$^{,}$\citep{Napiwotzki20}$^{,}$\citep{Kupfer22}.
With the observational constraints, we investigated the  {possible formation pathway (see Figure\,~\ref{fig:fate}(b))} of this system by performing numerical simulations under the assumptions of solar metallicity and ``Dutch" mass-loss scheme\citep{2009A&A...497..255G} with an efficiency of $\eta = 0.50$ (Methods).
 {The nature of short period suggests that Lan\,11 is formed through two phases of common envelope (CE) ejection processes. 
The primordial binary is initially consisted of a $\sim 7-12$\,$M_\odot$ intermediate-massive star and a $\sim 2 - 3$\,$M_\odot$ main-sequence (MS) star. 
The  {more} massive one evolves first and fills its Roche lobe at 
the giant branch (RGB/AGB) with a He-burning core about $2.0-2.5$\,$M_\odot$. 
The binary then enters into the first CE phase given the large mass-ratio \citep{Ge2020}. 
The envelope is ejected due to the low binding energy, a binary consisted of a He star with mass around $2.0-2.5$\,$M_\odot$ and an evolved companion star is then formed. 
The radius of the 
He star can further expand to a few hundreds solar radius so as to fill its Roche lobe that leads to the so-called case BB mass transfer\citep{1976A&A....47..231S,Tauris2015}.
After this mass transfer, a massive ONe WD around 1.2\,$M_{\odot}$ (see Figure\,2 of ref.\citep{Tauris2015}) and a companion star with mass around $2.6-3.2M_\odot$ are then left. 
This system will enter the second CE phase when the companion star fills its Roche lobe.
After the ejection of the second CE, the Lan\,11-like system consisted of a sdOB star and a massive ONe WD, is formed.
Under the aforementioned conditions, our simulations successfully reproduce the Lan\,11 system with the mass of sdOB star ranging from $0.51$ to $0.64$\,$M_{\odot}$ (similar to the observed interval for Lan\,11).
SdOB star in such a mass range can cross the observed intervals of Lan\,11 very well in $T_{\rm eff}$--log\,$g$ diagram (see Supplementary Figure~\ref{fig:track}).
According to our simulations, this sdOB star will soon (after few ten Myrs) evolve to white dwarf, and this binary will become a double-degenerate system.
Due to gravitational wave radiation, this double-degenerate system will eventually merge in $500$ to $540$ Myr.
In a widely accepted scenario, the super-Chandrasekhar nature and the presence of a massive ONe WD in this binary strongly suggest that the final violent merger will trigger the electron-capture reactions involving $^{20}$Ne and $^{24}$Mg, then the ONe WD collapses to a NS -- the so-called AIC process (refs.\citep{Miyaji80}$^{,}$\citep{Nomoto84}).
This system may also explode as a type Ia supernova, if a detonation can be triggered in ONe WD matter\citep{Marquardt2015}.
This may be possible but has not yet been proven to work\citep{2014ApJ...785...61S}.}

\section*{Methods}
\textbf{Observations and data reduction.}
Lan\,11 (TIC\,79323477) was visited by TESS on September 15,  2021 to  December 02, 2021 (Sectors 43-45), with a cadence of two-minute.
The data was downloaded from the Mikulski Archive of Space Telescopes (MAST, \url{https://mast.stsci.edu/portal/Mashup/Clients/Mast/Portal.html}), which was properly reduced and calibrated by the TESS Science Processing Operations Center (SPOC) pipeline.
We adopted the Pre-search Data Conditioning Simple Aperture Photometry (PDCSAP) flux data to construct our light curve.
We note that the PDCSAP performed simple aperture photometry by removing the instrumental trends and the aperture contributions of neighboring stars identified in a pre-search data condition (PDC). 
The later function in PDCSAP is particularly important for Lan\,11 since the existences of two bright nearby stars with Gaia RP-band magnitudes of 13.97 and 14.35 just around 28$''$ away.
We carefully check the two neighbors in Zwicky Transient Facility (ZTF) database\cite{Bellm2019} and find their magnitudes are almost stable during 1500 days observations. 
 {To further check the robustness of the contamination corrections from neighboring stars, we compare phase-folded light curves (with period taken from Table\,1) in each sector (see Supplementary Figure~\ref{fig:lc_tess_sigle_sector}) and find they are consistent with each other very well.
Moreover, the ZTF $g-$ and $r-$band phase-folded light curves of Lan\,11 are also in excellent agreement with that of TESS within measurement errors.}
To this degree, the light curve amplitude should be properly corrected by the PDCSAP algorithm.

 {As shown in Supplementary Table\,1, a total of 34 medium resolution ($R \sim 7500$) spectra were taken by LAMOST during October 2017 and February 2021.
The typical exposure time is from 10 to 20 min, yielding spectra with typical SNR around 5 in the blue arm and 10 in the red arm.
The high-quality time-series} optical spectra were obtained by DBSP installed on Palomar 200-inch telescope.
In total, twenty-eight exposures of 600s were taken with a slit width of 0.\arcsec5 on February 19, 2022 (see Supplementary Table\,2).
Unfortunately, the red arc lamp is broken during that night.
But the spectra at blue arm 
are well-observed with three exposures of FeAr arc taken during the scientific observations.
Given a resolving power of 2500, the obtained spectra are sufficient to derive accurate radial velocities, as well as the determinations of stellar parameters for Lan\,11.
The standard data reductions, including bias subtraction, flat correction, cosmic ray removal, one-dimensional-spectrum extraction, wavelength calibration, and flux calibration were performed 
\lijiao{by using a Python package {\tt pyexspec} developed ourselves}.

\textbf{FAST observation and data reduction.}
We performed a total of 2,300s of targeted exceptionally sensitive radio follow-up observation for Lan\,11 using the Five-hundred-meter Aperture Spherical radio Telescope (FAST) on Dec 05, 2022.
Note the first 1 min and the last 1 min were used for signal injection and polarization calibration.
The  Observation taken at FAST are using the center beam of the 19-beams L-band receiver, that the frequency range is from 1.05 to 1.45 GHz with an averaged system temperature around 25 K (ref.\citep{Jiang20}). 
Observation data is recorded in pulsar search mode, stored in PSRFITS format\citep{Hotan04}. 
Two types of data processing were performed for this observation: dedicated and blind search and single pulse search.

{\it Dedicated and blind search.}
Due to model dependence of the Galactic electron density model NE2001 (ref.\citep{CL02}$^{,}$\citep{Yao17}) and for the sake of robustness, we created de-dispersed time series for each pseudo-pointing over a wide range of dispersion measures (DMs), from 0 to 2,000 pc\,cm$^{-3}$, which is a factor of 7 larger than the Galactic maximum DM = 272.56 pc cm$^{-3}$ that models predicted in this line of sight. 
For each of the trial DMs, we searched for a periodical signal and first two order acceleration in the power spectrum based on the PRESTO pipeline\citep{Ransom01}$^{,}$\citep{Wang21}. We checked all the pulsar candidates with SNR greater than 6 pulse-by-pulse and removed the narrow-band radio frequency interferences (RFIs).

Both the periodical radio pulsations and single-pulse blind searches were performed for each observing epoch, but resulted in non-detections for all searches. 
We calibrated the noise level of the baseline, and then measured the amount of pulsed flux above the baseline, giving a 6$\sigma$ upper limit of flux density measurement of 41 $\pm$ 3 $\mu$Jy for persistent radio pulsations (assuming a pulse duty cycle of 0.05 to 0.3). 

{\it Single pulse search}
The above search strategy was continued to apply to de-disperse the time series for single pulse search. 
The soft packages PRESTO and HEIMDALL (\url{https://sourceforge.net/p/heimdall-astro/wiki/Home/}) are adopted.
A zero-DM matched filter was applied to mitigate RFI in the blind search. All of the possible pulse-candidates were plotted, then be checked as RFIs by manual eyes one-by-one. 
No pulsed radio emission with dispersive signature was detected with SNR $> 6$. 
The upper limit of pulsed radio emission is $\sim$0.15 Jy ms by assuming a 1 ms wide burst in terms of integrated flux.

\textbf {Stellar properties and orbital parameters of Lan\,11}
 {We present an overview of the estimates of stellar properties and orbital parameters in Supplementary Figure~\ref{fig:flowchart}, by using the observation data described above. The estimates include period determination, radial velocity (RV) curve fitting, spectral template matching, spectral energy distribution (SED) fitting, and light curve fitting, which will be detailedly  described in the following subsections one-by-one. The main results are summarized in Table\,1.}

\textbf{Period determination}
We performed a Lomb-Scargle periodogram for the TESS light curve using \lijiao{the function of {\tt LombScargle} implemented in} {\tt Astropy}\citep{astropy2018}. \lijiao{To search realistic periodicity, the frequency range is set from $\frac{1}{T_{\rm base}}$ to $\frac{1}{2\Delta T}$ with a step of $0.2 \times \frac{1}{T_{\rm base}}$, where $T_{\rm base} = 76.457$ days and $\Delta T = 120$\,s are the total observation duration (three sectors here) and the cadence of TESS light curve, respectively.}
The periodogram shows a significant peak around 110 min (see Supplementary Figure~\ref{fig:lc_period}).
We apply a Gaussian fit to the periodogram and find the peak value of 109.54440(20) min.
Phase-folded light curve to twice the peak value shows a clear Doppler beaming effect in the consecutive maxima with a height difference \zcj{around 0.2\%} (see Figure\,1(a)), which is consistent with the amplitude of the radial velocity curve (see Figure\,1(b)).
This result suggests that the orbital period of Lan\,11 is twice of the peak value, i.e. $P=219.08880(40)$ min,  {which is also clearly detected in the periodogram as the second harmonic (see Supplementary Figure~\ref{fig:lc_period})}.
The light curve of Lan\,11 shows a semi-amplitude of 1\% ellipsoidal variations, which is caused by the strong tidal force of its companion.

\textbf{Radial velocity curve and fitting}
Lan\,11 was first identified as a variable stars from its radial velocity measured from the LAMOST medium resolution survey.
However, due to the short wavelength coverage, low spectral SNR, and long exposure time (10-20 min), the high-quality time-series spectra were further collected using DBSP equipped in P200 telescope. 

{\it Radial velocity from LAMOST}
We measured the radial velocity from H$\alpha$ lines in the red arm {\color{black} spectra} by fitting a Gaussian function (for the absorption line) plus a second-order polynomial (for the continuum) with the IDL software package {\tt mpfit}.
Other lines were not used for radial velocities either due to the low SNR in the blue arm or the weak lines in the red range.
The typical uncertainties of the measurements are 10\,km\,s$^{-1}$.
We applied a simple sinusoidal function to fit the radial velocity curve by fixing the orbital period found from the TESS light curve.
The semi-amplitude found by the fitting is $251.15 \pm 10.00$\,km\,s$^{-1}$.

{\it Radial velocity of DBSP spectra}
The radial velocities were measured from the spectra between  {3750} and 5200\,\AA\, obtained from P200 DBSP by using the cross-correlation function (CCF) method with a grid of synthetic spectra for subdwarfs. 
 {The grid includes 96 synthetic spectra with effective temperature ranging from 10,000 to 65,000\,K (in a step of 5000\,K for $T_{\rm eff} \le 45,000$\,K and a hottest template with $T_{\rm eff} = 65,000$\,K), surface gravity covering 4.5 to 6.5 (in a step of 1.0\,dex) and 
four representative chemical abundances.  
Please refer to ref.\cite{Pacheco2021} for details.}
A Gaussian kernel with FWHM\,$\sim 1.8$\,\AA\, was used to degrade the synthetic spectra to fit the DBSP resolution ($R \sim 2500$). 
 {We then derive the radial velocity for each DBSP spectrum by using the CCF method developed in ref.\cite{Zhangbo2021ApJS..256...14Z}, with the above degraded synthetic spectra and a radial velocity grid of $-500$ to $500$\,km\,s$^{-1}$ in a step of 5\,km\,s$^{-1}$.}
The uncertainties of the measured radial velocities were evaluated using the Monte Carlo (MC) method by sampling the spectra 100 times using the individual flux error of each pixel.
The whole measurements were realized by using the radial velocity module in {\tt Laspec}\cite{Zhangbo2020}.
The results are presented in Supplementary Table\,1.


{\it Radial velocity curve fitting}
By fixing the orbital period $P = 219.08880$ min obtained from the TESS light curve, we derived the orbital parameters of the Lan\,11 by fitting its radial velocity curve measured from DBSP spectra using the Markov Chain Monte Carlo (MCMC) approach. The likelihood of the parameters are adopted such that
\begin{equation}\label{eq:proprv}
   \begin{split}
    \ln[\mathcal{L}(\theta_v \mid v)] &\propto \ln [\mathcal{L}(v \mid \theta_v)] \\
    &= -\frac{1}{2}\sum_{i=1}^{N}\frac{[v_{i}-v(\theta_v, t_i)]^2}{\sigma^2_{v_i}+s^2}\\ &-\frac{1}{2}\sum_{i=1}^{N}\ln[2\pi(\sigma_{v_i}^2+s^2)],
  \end{split}
\end{equation}
where $v_{i}$ and $\sigma_{v_{i}}$ are the radial velocities and their measurement errors, respectively, {\color{black} measured from} the DBSP spectra, and $\theta_v = (P, T_0, K_{\rm sd}, v_\gamma, \sqrt{e}\cos\omega, \sqrt{e}\sin\omega)$ are the orbital parameters.
$K_{\rm sd}\ e,\ \omega$, and $v_\gamma$ are the radial velocity semi-amplitude, eccentricity, longitude of periastron, and the systematic radial velocity.
 $v(\theta_v, t_i)$ represents the Kepler orbital radial velocity at epoch $t_i$ with orbital parameters $\theta_v$, calculated by \lijiao{the function {\tt rv\_drive} implemented in} the python package {\tt radvel}\cite{Fulton2018}.
$s$ is an additional “jitter” variance which may absorb all kinds of modeling errors such as errors of the instrument, wavelength calibration, and even the radial velocity model. 
 {We note that the value of $v(\theta_v, t_i)$ is calculated by considering the exposure duration (10 min, about 5\% of the period):
\begin{equation}
    v(\theta_{v}, t_i) = \frac{1}{10 \, {\rm min}}\int_{t_i-5\, {\rm min}}^{t_i+5\, {\rm min}}{v(\theta_{v}, t) {\rm d} t}\text{,}
\end{equation}
where $t_{i}$ is the middle time of each exposure.}
The fitting results are listed in Table\,1,  {the posterior distributions of these orbital parameters are shown in Supplementary Figure~\ref{fig:rv_coner}}.
The semi-amplitude of the radial velocity curve from DBSP spectra is 
\lijiao{$K_{\rm sd} = 249.8\pm2.1$} \kms, which is consistent with the result found from LAMOST radial velocity measurements.
 {A small eccentricity $e = 0.02 \pm 0.01$ is found after the fitting procedure.
We note this is only an upper limit, given the 10 min exposure duration and sparse sampling of P200 spectroscopic observations.
The eccentricity will be well constrained by the high-precision and well-sampling TESS light curve in the later analysis.}

\textbf{Kinematics of Lan\,11}
We determined the 3D velocity of Lan\,11 by using the positions, distance, proper motions from {\it Gaia}, and the systemic velocity from the radial velocity curve fitting.
During the calculation, the Sun is placed at $(R, Z) = (8.178, 0.025)$\,kpc (refs.\citep{GRAVITY19}$^,$\citep{Bland-Hawthorn16}) with peculiar velocities with respect to the Local Standard of Rest $(U_{\odot}, V_{\odot}, W_{\odot}) = (7.01, 10.13, 4.95)$\,km\,s$^{-1}$ (ref.\citep{Huang15}).
The circular velocity at the solar position was assumed as $V_{\rm c} (R_0) = 234.04$\,km\,s$^{-1}$ (ref.\citep{Zhou23}).
The resulting 3D velocity of Lan\,11 is $(V_R, V_{\phi}, V_Z) = (-11.90^{+1.58}_{-1.47}, 237.80^{+0.18}_{-0.18}, -8.31^{+0.37}_{-0.34})$\,km\,s$^{-1}$.
The large azimuthal velocity implies that Lan\,11 is a thin disk star.
The Galactic orbits ({ see  Supplementary Figure~\ref{fig:3d_orbit}}) for Lan\,11 were further calculated by using the python package {\tt Gala}\citep{Price-Whelan17} with the potential setting to {\tt MilkyWayPotential}\citep{Bovy15}.
The orbits show that the eccentricity $e$ and the maximum distance above the Galactic plane $z_{\rm max}$ are 0.06 and 0.16\,kpc, respectively. This is again evident that Lan\,11 belongs to the thin disk population.

\textbf{Spectral template matching with {\sc XTgrid}}
 {The accurate determinations of atmospheric parameters of hot subdwarf in such a close binary is challenging. The choices of proper spectra should consider 1) the quality (SNR), 2) the phase (avoid effects of geometric distortions from the tidal force of the unseen object) and 3) the number of spectra (avoid line broadening from Doppler shift corrections).
Doing so, two choices of spectra are adopted to derive reliable atmospheric parameters of Lan 11.
First, the DBSP spectrum with the highest SNR ($> 135$) was chosen to meet requirements 1) and 3).
Second, four high quality (SNR $> 100$) DBSP spectra at different phases (see Supplementary Table\,2) were chosen to meet requirements 1) and 2).}
The atmospheric parameters of Lan\,11 were  {then} measured by matching  {either the highest SNR DBSP spectrum or four phase-picked high-SNR DBSP spectra} with the synthetic spectra using a steepest-descent iterative spectral analysis implemented in the {\sc XTgrid} code\cite{Nemeth19}.
The synthetic spectra were generated by the Non-Local Thermodynamic Equilibrium (NLTE) atmospheric models of {\tt TLUSTY} (v207) and {\tt SYNSPEC} (v53)\cite{Hubeny1995, Hubeny2017}.
By starting with an initial guess, {\sc XTgrid} applies a steepest-descent $\chi^2$ minimization, based on successive approximations, until the parameters converge to the best solution.
The procedure calculates new models on the fly and adjusts the model parameters and atomic data inputs to precisely link the variations in the theoretical atmospheric structure to the observable emergent spectrum.  
We applied H, He, C, N, O, Mg, and Si opacities in the model calculations and, in addition, S in the spectral synthesis. 
All comparisons were done globally, using the observed spectra with the wavelength range of  {$3750-5200$\,\AA\, (with Ca K line masked out to avoid the influence of interstellar medium absorption)} and a piecewise normalization of the model to the observation. 
During this process, the effective temperature, surface gravity, projected rotation velocity, and chemical abundances are adjusted independently to minimize the global $\chi^2$. 
{ As an example,} the best-fit model  {for the highest SNR DBSP spectrum} is shown as the red line in Figure~\ref{fig:spec-sed} (a). 
More comprehensive comparisons for the individual Balmer and helium lines are shown in Supplementary Figures~\ref{fig:dbsp_lines}--\ref{fig:dbsp_lines_phase_picked}.
During error calculations, new models are calculated radially outward from the best-fit point until the confidence limit is reached. 
 {The best-fit parameters for the highest SNR spectrum and four phase-picked high-SNR spectra are presented in Table\,1.
By combining the two groups of estimates, the $1\sigma$ interval is [35,710, 35,990]\,K for $T_{\rm eff}^{\rm sd}$, [5.29, 5.36] for $\log g_{\rm sd}$, and [$-0.97$, $-0.83$] for $\log n{\rm He}/n{\rm H}$.}


\textbf{Spectral energy distribution fitting}
We used the spectral energy distribution (SED) fitting method to determine the radius, luminosity and extinction of Lan\,11. 
 {A python program {\tt SPEEDYFIT}\footnote{https://speedyfit.readthedocs.io/en/stable/index.html} was adopted to determine the stellar parameters, as well as their uncertainties, by using the MCMC approach.}
During the fitting, photometric fluxes and their uncertainties from {\it Gaia} EDR3, PanSTARRS, 2MASS, and ALLWISE and parallax measurements from {\it Gaia} EDR3, as well as Gaussian priors on the effective temperature and log$g$ from the spectroscopic solutions, were applied.
The synthetic spectral model grid for hot subdwarf stars was generated by `TMAP', which is equipped by {\tt SPEEDYFIT}.
{The posterior probability distributions of these parameters are shown in Supplementary Figure~\ref{fig:SED_corner}.}
The best-fitted SED by a single hot subdwarf agrees very well with the observed multi-band photometry (see Figure~\ref{fig:spec-sed}(b)).
This result suggests the contribution to the observed SED from the unseen companion is negligible, {implying a compact nature of the dark companion.}
The resulting parameters are listed in Table\,1.

\textbf{Light curve fitting}
 {In principle, the ellipsoidal light curve of such a close binary is mainly determined by the orbital parameters and stellar properties ($T_{\rm eff}^{\rm sd}$, log\,$g_{\rm sd}$ and $R_{\rm sd}$) of the visible star.
The mass function of Lan\,11 has been determined from the RV curve fitting, and the stellar properties are well constrained from the spectral template matching and the SED fitting.
With these known parameters as priors (see Supplementary  Figure~\ref{fig:flowchart}), the binary inclination angle $i$ and eccentricity $e$ can be well extracted from the light curve.}
The TESS light curve is modelled with two popular packages: {\tt ellc}\citep{Maxted16} and {\tt PHOEBE}\citep{PZ05}, respectively.
Before modeling, the TESS light curve from Sectors 43-45 is normalized by using a python package {\tt LightKurve}\cite{lightkurve} and significant outliers were removed by a $3\sigma$-clipping procedure.
The normalized light curve was then phase-folded and further divided into 50 equal bins.
We then derived the mean value of the flux at each phase bin, as well as the uncertainty of the mean properly estimated from error propagation.
The final average light curve is presented in Figure~\ref{fig:lc-rv} (a). 
Typically, the uncertainty of light curve at each phase bin is 300 ppm, which is in consistent with the independent estimate from the bootstrap technique.
 {Here the function {\tt bootstrap} implemented in {\tt scipy} was used.  We adopted the default settings  except for changing {\tt confidence level = 0.68} to estimate the $1\sigma$ error of the mean value. }


As mentioned earlier, the Doppler beaming signal is clearly detected in the phase-folded light curve, which should also be carefully considered in the light curve modeling. 
Under the assumption of radial velocity much smaller than light speed, the beam effect can be expresses as,
\begin{equation}
    F_{\lambda} = F_{0, \lambda}\left(1-B\frac{v}{c}\right),
    \label{eq:beaming}
\end{equation}
where $F_{\lambda}$ and $F_{0, \lambda}$ represent the observed flux and emitted flux, respectively.
Here $v$ and $c$ are radial velocity and speed of light.
$B$ is the beaming factor that is dependent on the wavefront and the relativistic Doppler shift of the target spectrum (with extinction considered) within the TESS photometric band.
By using the NLTE synthetic spectrum with the exact parameters from our spectral template matching, {we compute synthetic photometric flux at different radial \zcj{velocities} and find} the beaming factor $B$ to be \zcj{1.2034} for Lan\,11.

{\it Light curve fitting with {\tt ellc}} 
Since the companion should be a compact object, the surface brightness ratio is set to 0 and $R = 0.01 R_{\odot}$, providing the companion is a WD.
During the light curve modeling, the limb and gravity darkening (with coefficients interpolated from the tables provided by ref.\citep{Claret2020AA}) and the Doppler beaming effects ($B = \zcj{1.2034}$) were properly considered.
The reflection was ignored.
 {In the model, period was fixed to the solution from {\tt LombScargle}}, and 
$T_{\rm eff}^{\rm sd}$, $\log g_{\rm sd}$, $R_{\rm sd}$ and $K_{\rm sd}$ were assumed to be Gaussian distributions with solutions and standard deviations from the spectrum template matching, the SED fitting, and the radial velocity curve fitting.  
 {An upper limit $e =0.02$ yielded by radial velocity curve fitting was also considered. The full settings are summarized in Supplementary Table\,2.}
Finally, the model includes {four} free parameters: the orbital inclination $i$ and the superior conjunction time $T_0$, { $\sqrt{e}\cos\omega$ and $\sqrt{e}\sin\omega$}.
With above parameter setups in {\tt ellc}, an MCMC approach was performed to sample the light curve by using {\tt EMCEE}\cite{emcee2013}).
 {Note that two groups of parameters are yielded by the fitting since the priors of atmospheric parameters ($T_{\rm eff}^{\rm sd}$ and $\log g_{\rm sd}$) can be taken either from the best SNR spectrum or from four phase-picked spectra.
As an example, the posterior probability distributions of the parameters with priors from the highest SNR spectrum are shown in Supplementary Figure~\ref{fig:ellc_coner}.}
 
{\it Light curve fitting with PHOEBE} 
With the similar parameter setups, the light curve was modeled independently by {\tt PHOEBE}\citep{PZ05}, which is an open-source python-based module to compute theoretical light and radial velocity curves. 
We note the gravity darkening is calculated by {\tt PHOEBE} itself \zcj{with the bolometric gravity brightening coefficient fixed to 1}.
The coefficients of limb-darkening is interpolated by a logarithmic relation in {\tt PHOEBE} rather than the four-term relation adopted in {\tt ellc}.
 {Again, the posterior probability distributions of the parameters with priors from highest SNR  spectrum are shown in Supplementary Figure~\ref{fig:phoebe_coner} for an example}.

For both {\it ellc} and {\tt PHOEBE}, we ran \zcj{27,000} steps in MCMC.
The first 2,000 steps were ignored as burn in.
The best-fit results are summarized in  {Supplementary  Table\,3.}
 {The final adopted values of the orbital and physical parameters of this system were set to an interval ranging from the minimum values minus its $1\sigma$ errors to the maximum values plus its $1\sigma$ errors from the four groups of estimates determined by the two softwares.
The results are listed in Table\,1.} 
The model light curves reconstructed by using the best-fit parameters { (with priors from the highest SNR spectrum)} from {\tt ellc} and {\tt PHOEBE} are shown in Figure\,1(a).


{ \textbf{The possible nature of the companion star by Binary population synthesis}}
 {The companion of Lan\,11 is strongly suggested to be a compact star by the well-fitted SED from a single hot subdwarf and the non-detection of double-line signal in all of the collected spectra. 
Here we further explore the nature of this compact object: a CO WD, an ONe WD, or even a NS?
Doing so, we performed Monte Carlo binary population synthesis (BPS) calculations to generate WD/NS+hot subdwarf binary populations\citep{Liu20}, by employing the Hurley rapid binary evolution code\citep{HUR00}$^{,}$\citep{HUR02}. 
The basic assumptions and initial parameters for the Monte Carlo BPS calculations are:
(1) All stars are assumed to be members of binaries with circular orbits.
(2) The initial mass function of primary stars in the primordial binaries is assumed to follow that from ref.\citep{mil79}, and the mass ratio \zcj{to follow uniform distribution}.
(3) The initial distribution of orbital separation $a$ is taken from  ref.\citep{HAN95} that contains a constant value in $\log\,a$ for the wide binaries (orbital period larger than 100\,yr) and a smoothly falling off distribution for close binaries (orbital period smaller than 100\,yr).
This distribution ensures a roughly equal number of binaries in both wide and close orbits.
(4) The common envelope (CE) ejection process plays a very important role in the formation of WD/NS+hot subdwarf binary systems.
We adopted the standard energy prescription to deal with the CE ejection process\citep{Ivanova13} by using two free parameters, i.e. the CE ejection parameter $\alpha_{\rm CE}$ and the stellar structure parameter $\lambda$.
The setting of $\lambda$ is same as refs.\citep{Loveridge11}$^{,}$\citep{ZL14}.
For the value of $\alpha_{\rm CE}$, we tried three groups of values, i.e. 0.5, 0.75, and 1.0, which cover the full range of common adopted values.
For each choice of $\alpha_{\rm CE}$, $2\times10^8$ primordial binaries are generated and evolved from their births to the formation of WD/NS$+$hot subdwarf binary systems within Hubble time.}

 {In each of our simulations, LAN\,11-like systems are identified by requiring their masses and orbital periods similar to these of Lan\,11, i.e., $0.52 \le M_{\rm sd} \le 0.65$\,$M_{\odot}$, $1.08 \le M_{\rm comp} \le 1.35$\,$M_{\odot}$, $|P_{\rm BPS} - P_{\rm Lan 11}| \le 200$\,min (see Table\,1).
The nature of the compact star in our BPS simulation is decided by its degenerate CO core mass of its progenitor at the AGB or He giant stage.
It is classified as a CO WD if the core mass smaller than $1.088$\,$M_{\odot}$, or as an ONe WD if the core mass ranging from $1.088$\,$M_{\odot}$ to the Chandrasekhar mass limit, or as a NS if the core mass greater than the Chandrasekhar mass limit but less than $6.5\,\rm M_{\odot}$ (refs.\citep{Tauris2015}\citep{Lauffer18}).
The possibilities of LAN\,11 like systems with a CO WD, an ONe WD or a NS are then calculated and listed in Supplementary Table\,4.
In all cases, the probability of ONe WD found in Lan\,11-like systems is at least 85\%, and the total probability for CO WD and NS is no more than 15\%.
In this regard, the nature of the compact companion of Lan\,11 is most likely an ONe WD.}

\textbf{Evolution of the system}
The formation and evolutionary models of LAN 11 were investigated via the Modules for Experiments in Stellar Astrophysics (MESA; release 12115)\citep{Paxton11}.
We applied the ``Dutch'' wind prescription\citep{2009A&A...497..255G} with efficiency $\eta=0.5$ for the wind mass-loss. 
The metallicity is set to be solar value ($Z = 0.02$). 
The parameters of mixing length, semi-convective, overshooting are the same as in ref.\citep{Sen22}. 
In the standard formation channel of hot subdwarf stars\citep{Han03}, the short orbital period of Lan\,11 suggests that its progenitor binary has undergone two phases of common envelope ejection processes. 
During the CE phase, the star is unable to accrete much of material due to the extremely short timescale\citep{Ivanova13}. 
 {After the first CE, there is enough time for the He star to expand to fill its Roche lobe before the start of second CE.
The companion will accrete mass from this expending He star, which is the so-called Case BB mass transfer\citet{1976A&A....47..231S,Tauris2015}. 
The mass of the the formed WD is mainly decided by the mass of the He star (see Figure\,2 of ref.\citep{Tauris2015}).}


We then tried to find the progenitor of sdOB star in Lan\,11 with a grid of hot subdwarf models. 
The models are constructed in a way similar to that of ref.\citep{HC13}, but with masses of progenitors in the range of $2.5  - 3.5\,M_\odot$. 
First, the binary enters CE phase when the second one evolved to near the tip of RGB.
The envelope was then removed with a high mass-loss rate ($\sim 10^{-3}M_\odot\;\rm yr^{-1}$).
{ For each progenitor, the remnant mass is fixed in four cases, i.e. $0.51$, $0.55$, $0.59$, and $0.64$ $M_{\odot}$, which covers the possible observable interval of Lan\,11.
Then the hydrogen envelope mass and helium core mass are changed accordingly. 
By comparisons between the evolutionary tracks and observations (see Supplementary Figure~\ref{fig:track}), the possible progenitor masses for the hot subdwarf of $0.51$, $0.55$, $0.59$, and $0.64$ $M_\odot$ are $2.6$, $2.8$, $3.0$, and $3.2$\,$M_{\odot}$, respectively.} 
For each sdOB star, if a low-mass progenitor ($\lesssim 2.6$\,$M_\odot$) is adopted, the helium core at the RGB is so small that we cannot construct the hot subdwarf models with the corresponding mass. 
On the other hand, if the progenitor is more massive ($\gtrsim 3.2$\,$M_\odot$), none of the evolutionary tracks can match the observations in the log\,$T_{\rm eff}$--log$g$ diagram. 
The main reason is that the hot subdwarf produced from a massive progenitor generally has a dense envelope due to the hydrogen burning in the shell, which leads to a large value of $\log g$ (small radius) compared to the observations (see also ref.\cite{Kupfer2020}).

{ 
The four best-fit models with sdOB mass of $0.51$, $0.55$, $0.59$, and $0.64$ $M_\odot$ are shown in Supplementary Figure~\ref{fig:track}(a). 
The tracks start from He core burning with corresponding hydrogen envelope mass of $0.026$, $0.030$, $0.033$, and $0.037$\,$M_\odot$, respectively. 
The relatively large envelope masses are required since the small value of $\log g$ (corresponds to a large radius) of sdOB in Lan\,11, as compared to the majority in observations typically resulted from standard thin and light envelope (grey dots in Supplementary Figure~\ref{fig:track}(a)).
For the case of $M_{\rm sd}\geq 0.55M_\odot$, the sdOB star will fill its Roche lobe due to the shell burning, which leads to the mass transferring to the compact object (see below for detailed explanations). On the contrary, the mass transfer phase will not be initiated for the case of $M_{\rm sd} < 0.55M_\odot$, where the sdOB radius is always within its Roche lobe.

Finally, we present a representative example of the evolutionary details with $M_{\rm sd}=0.59M_\odot$ (corresponding to progenitor mass of 2.8\,$M_\odot$) and compact object mass of $1.29M_\odot$ in Supplementary Figure~\ref{fig:track}(b). 
The track starts from He core burning with envelope mass of $0.034\,M_\odot$, and terminates at the merger event. 
The He core burning sustains for $\sim 7\times 10^7\;\rm yr$ (from stage 1 to 2 as shown in the inset of Supplementary Figure~\ref{fig:track}(b)). 
Then the residual hydrogen burning in the envelope leads to the expansion of sdOB radius (from stage 2 to 3), where the current Lan\,11 locates between stage 2 and 3. 
In the subsequent evolutions, we found that the hot subdwarf will undergo two mass transfer phases after He core burning. 
One is supported by the H burning in the shell (from stage 3 to 4), and another is supported by the He burning in the shell (from stage 6 to 7). 
After 500-540 Myr, the binary will merge and may end up as an AIC\citep{Miyaji80}$^{,}$\citep{Nomoto84}, although the possibility of SNe Ia cannot be fully ruled out\citep{Marquardt2015}. 
The whole evolution path of this system is summarized in Fig\,3(b).}

\section*{Data Availability}
The data supporting the plots in this paper
and other findings of this study are available from the corresponding authors upon reasonable request.

\setcounter{figure}{0}
\renewcommand{\figurename}{Supplementary Figure}

\begin{table*}
\small{\textbf{Supplementary Table 1: LAMOST Observation Log for Lan\,11.}  {For the sake of simplicity, the phase is calculated by fixing $T_0 = 2459477.82316$ derived from radial velocity curve fitting.}}
\small
\centering
\begin{tabular}{ccccccc}
\hline\hline
Facility     & UT shut   & BJD\footnotesize{mid} & Exposure & RV & Phase &SNR\\
  &     yyyy-mm-dd hh:mm:ss &day&s&${\rm kms^{-1}}$  &  n.a.&n.a.\\
\hline

\multirow{34}{*}{LAMOST}
& 2017-10-29T20:18:00 & 2458056.353515 & 600 &$ -242.61 \pm 6.48 $&0.139& 16.61 \\
& 2017-10-29T20:31:00 & 2458056.362543 & 600 &$ -243.15 \pm 8.05 $&0.198& 14.22 \\
& 2017-10-29T20:45:00 & 2458056.372266 & 600 &$ -276.79 \pm 9.53 $&0.262& 14.20 \\
& 2017-10-29T20:58:00 & 2458056.381295 & 600 &$ -204.23 \pm 7.56 $&0.321& 12.75 \\
& 2017-10-29T21:15:00 & 2458056.393101 & 600 &$ -144.62 \pm 9.59 $&0.399& 14.23 \\
& 2017-10-30T20:23:00 & 2458057.357066 & 600 &$ 233.72 \pm 6.52 $&0.735& 16.97 \\
& 2017-10-30T20:37:00 & 2458057.366789 & 600 &$ 201.54 \pm 8.47 $&0.799& 14.96 \\
& 2017-10-30T20:50:00 & 2458057.375818 & 600 &$ 161.74 \pm 7.16 $&0.858& 18.00 \\
& 2017-10-30T21:04:00 & 2458057.385541 & 600 &$ 97.08 \pm 6.43 $&0.922& 17.63 \\
& 2017-11-02T19:57:00 & 2458060.339239 & 600 &$ -216.32 \pm 11.00 $&0.335& 13.78 \\
& 2017-11-02T20:16:00 & 2458060.352435 & 600 &$ -101.51 \pm 11.33 $&0.422& 12.16 \\
& 2017-11-02T20:30:00 & 2458060.362158 & 600 &$ 16.25 \pm 11.64 $&0.486& 12.89 \\
& 2017-11-04T18:18:00 & 2458062.270632 & 600 &$ -100.77 \pm 14.43 $&0.030& 6.18 \\
& 2017-11-04T18:32:00 & 2458062.280355 & 600 &$ -197.21 \pm 18.24 $&0.094& 6.68 \\
& 2017-11-04T18:45:00 & 2458062.289384 & 600 &$ -233.37 \pm 12.19 $&0.153& 6.70 \\
& 2019-02-21T11:15:00 & 2458535.979192 & 1200 &$ 81.84 \pm 17.51 $&0.563& 8.55 \\
& 2019-02-21T11:39:00 & 2458535.995858 & 1200 &$ 229.07 \pm 17.51 $&0.673& 10.95 \\
& 2019-02-21T12:02:00 & 2458536.011828 & 1200 &$ 258.95 \pm 24.49 $&0.778& 6.80 \\
& 2019-02-21T12:26:00 & 2458536.028494 & 1200 &$ 160.72 \pm 15.63 $&0.888& 10.29 \\
& 2019-02-21T12:49:00 & 2458536.044464 & 1200 &$ 24.50 \pm 19.07 $&0.992& 6.53 \\
& 2019-02-21T13:12:00 & 2458536.060435 & 1200 &$ -189.92 \pm 20.18 $&0.097& 6.70 \\
& 2019-12-05T17:06:00 & 2458823.225701 & 1200 &$ 155.05 \pm 16.86 $&0.542& 11.92 \\
& 2019-12-05T17:35:00 & 2458823.245840 & 1200 &$ 227.73 \pm 10.27 $&0.674& 13.28 \\
& 2019-12-05T17:58:00 & 2458823.261813 & 1200 &$ 268.37 \pm 18.21 $&0.779& 12.27 \\
& 2021-01-20T12:41:00 & 2459235.041135 & 1200 &$ -296.58 \pm 17.39 $&0.272& 11.01 \\
& 2021-01-20T13:04:00 & 2459235.057106 & 1200 &$ -162.81 \pm 14.80 $&0.377& 9.32 \\
& 2021-01-20T13:28:00 & 2459235.073772 & 1200 &$ -15.29 \pm 16.11 $&0.487& 9.15 \\
& 2021-01-30T12:40:00 & 2459245.039865 & 1200 &$ -39.25 \pm 23.59 $&0.991& 8.38 \\
& 2021-01-30T13:04:00 & 2459245.056530 & 1200 &$ -187.26 \pm 10.68 $&0.100& 9.80 \\
& 2021-01-30T13:27:00 & 2459245.072501 & 1200 &$ -247.41 \pm 10.71 $&0.205& 11.63 \\
& 2021-02-02T13:57:00 & 2459248.093134 & 1200 &$ -107.73 \pm 8.20 $&0.059& 14.11 \\
& 2021-02-02T14:21:00 & 2459248.109799 & 1200 &$ -241.51 \pm 10.39 $&0.168& 13.23 \\
& 2021-02-02T14:47:00 & 2459248.127854 & 1200 &$ -214.24 \pm 9.60 $&0.287& 13.02 \\
& 2021-02-20T11:15:00 & 2459265.979227 & 1200 &$ 185.05 \pm 19.69 $&0.618& 8.13 \\
\hline\hline
\end{tabular}
\label{tab:rv_lamost}
\end{table*}

\begin{table*}
\small{\textbf{Supplementary Table 2: DBSP Observation Log for Lan\,11.}  {For the sake of simplicity, the phase is calculated by fixing $T_0 = 2459477.82316$ derived from fitting radial velocity curve. The four spectra marked with $*$ are selected for spectral fits.}}
\small
\centering
\begin{tabular}{ccccccc}
\hline\hline
Facility     & UT shut   & BJD\footnotesize{mid} & Exposure & RV & Phase &SNR\\
  &     yyyy-mm-dd hh:mm:ss &day&s&${\rm kms^{-1}}$  &  n.a.&n.a.\\
\hline
\multirow{28}{*}{P200 DBSP}
& 2022-02-19T03:35:50$^*$ & 2459629.657012 & 600 & $78.29\pm0.97$ & 0.955 & 137.79\\
& 2022-02-19T03:59:21$^*$ & 2459629.673343 & 600 & $-98.39\pm1.39$ & 0.062 & 110.51\\
& 2022-02-19T04:14:19 & 2459629.683740 & 600 & $-208.10\pm0.78$ & 0.131 & 105.52\\
& 2022-02-19T04:24:41 & 2459629.690935 & 600 & $-244.49\pm0.91$ & 0.178 & 102.13\\
& 2022-02-19T04:35:02 & 2459629.698130 & 600 & $-261.65\pm1.32$ & 0.225 & 100.98\\
& 2022-02-19T04:45:24 & 2459629.705325 & 600 & $-255.03\pm1.05$ & 0.272 & 100.91\\
& 2022-02-19T04:55:46$^*$ & 2459629.712520 & 600 & $-230.00\pm0.84$ & 0.320 & 108.16\\
& 2022-02-19T05:06:08 & 2459629.719714 & 600 & $-183.23\pm1.07$ & 0.367 & 103.04\\
& 2022-02-19T05:16:29 & 2459629.726909 & 600 & $-121.69\pm1.26$ & 0.414 & 105.22\\
& 2022-02-19T05:26:51$^*$ & 2459629.734104 & 600 & $-50.00\pm1.06$ & 0.462 & 110.40\\
& 2022-02-19T05:37:13 & 2459629.741299 & 600 & $20.23\pm0.78$ & 0.509 & 107.77\\
& 2022-02-19T05:47:34 & 2459629.748493 & 600 & $98.65\pm1.42$ & 0.556 & 104.29\\
& 2022-02-19T05:57:56 & 2459629.755688 & 600 & $155.28\pm1.20$ & 0.603 & 93.76\\
& 2022-02-19T06:19:17 & 2459629.770510 & 600 & $236.04\pm1.29$ & 0.701 & 83.41\\
& 2022-02-19T06:29:38 & 2459629.777704 & 600 & $241.38\pm1.04$ & 0.748 & 94.08\\
& 2022-02-19T06:40:00 & 2459629.784899 & 600 & $221.57\pm1.52$ & 0.795 & 93.00\\
& 2022-02-19T06:50:22 & 2459629.792094 & 600 & $187.36\pm1.10$ & 0.843 & 94.79\\
& 2022-02-19T07:00:43 & 2459629.799289 & 600 & $138.03\pm1.80$ & 0.890 & 92.90\\
& 2022-02-19T07:11:05 & 2459629.806484 & 600 & $85.85\pm0.91$ & 0.937 & 96.11\\
& 2022-02-19T07:21:27 & 2459629.813679 & 600 & $11.39\pm1.02$ & 0.985 & 94.41\\
& 2022-02-19T07:31:48 & 2459629.820873 & 600 & $-62.87\pm1.39$ & 0.032 & 92.46\\
& 2022-02-19T07:42:10 & 2459629.828068 & 600 & $-133.19\pm1.06$ & 0.079 & 90.19\\
& 2022-02-19T08:02:06 & 2459629.841904 & 600 & $-230.00\pm1.04$ & 0.170 & 84.26\\
& 2022-02-19T08:12:27 & 2459629.849099 & 600 & $-253.75\pm1.49$ & 0.217 & 85.53\\
& 2022-02-19T08:22:49 & 2459629.856294 & 600 & $-255.63\pm1.45$ & 0.265 & 82.68\\
& 2022-02-19T08:33:11 & 2459629.863489 & 600 & $-223.88\pm1.49$ & 0.312 & 73.21\\
& 2022-02-19T08:43:32 & 2459629.870684 & 600 & $-180.00\pm1.28$ & 0.359 & 70.06\\
& 2022-02-19T08:53:54 & 2459629.877878 & 600 & $-124.83\pm1.98$ & 0.407 & 68.19\\
\hline\hline
\end{tabular}
\label{tab:rv_p200}
\end{table*}

\begin{table*}[ht!]
\small{\textbf{Supplementary Table 3: Results from light curve fitting based on different software packages and priors.} $\mathcal{N}$ and $\mathcal{U}$ represent normal and uniform distributions, respectively. Quoted values are the median, and uncertainties denote the 68\% confidence interval.}
\centering 
\begin{tabular}{c c |c c |c c } 
\hline\hline  
\multirow{8}{*}{Prior}
 & $P$ (min) & \multicolumn{4}{c}{219.08880 (fixed)}\\
 &$K_{\rm sd}$ & \multicolumn{4}{c}{$\mathcal{N}(249.8, 2.1)$}\\
&$\sqrt{e}\cos\omega$& \multicolumn{4}{c}{$\mathcal{U}(-0.10, 0.10)$} \\
&$\sqrt{e}\sin\omega$& \multicolumn{4}{c}{$\mathcal{U}(-0.10, 0.10)$} \\
&$T_0$ (BJD-24572477)& \multicolumn{4}{c}{$\mathcal{U}(0.81, 0.83)$}\\
 &$\cos i$& \multicolumn{4}{c}{$\mathcal{U}(0, 1)$}\\
 \cline{3-6}
 &$T^{\rm sd}_{\rm eff}$& \multicolumn{2}{c|}{$\mathcal{N}(35840, 130)$} & \multicolumn{2}{c}{$\mathcal{N}(35850, 140)$}\\
 &$\log g_{\rm sd}$ & \multicolumn{2}{c|}{$\mathcal{N}(5.306, 0.02)$} & \multicolumn{2}{c}{$\mathcal{N}(5.338, 0.02)$} \\
 &$R_{\rm sd}$& \multicolumn{2}{c|}{$\mathcal{N}(0.275, 0.007)$} & \multicolumn{2}{c}{$\mathcal{N}(0.275, 0.007)$}\\
   \hline
   \hline
  & & {\tt ellc} & {\tt PHOEBE} & {\tt ellc} & {\tt PHOEBE} \\
   \hline 
\multirow{8}{*}{Posterior}
&$K_{\rm sd}$ (\kms) & $249.9_{-2.1}^{+2.1}$ &$249.7_{-2.1}^{+2.0}$ & $250.1_{-2.1}^{+2.1}$ & $249.7_{-2.1}^{+2.0}$\\
&$\sqrt{e}\cos\omega$ &$-0.029_{-0.011}^{+0.014}$ & $-0.056_{-0.014}^{+ 0.016}$& $-0.029_{-0.011}^{+0.014}$ & $-0.056_{-0.014}^{+0.016}$ \\
&$\sqrt{e}\sin\omega$ & $-0.004_{-0.019}^{+0.019}$  & $0.014_{-0.022}^{+0.022}$& $-0.003_{-0.019}^{+0.018}$ &$0.016_{-0.023}^{+0.022}$\\
& $T_0$ (BJD-2459477) & 0.82172(11) & 0.82182(13)& 0.82172(11) & 0.82182(13) \\
&$i$ (\degree) & $48.3_{-1.6}^{+2.0}$ &$47.9_{-2.5}^{+2.5}$&$50.8_{-2.4}^{+2.8}$ &$52.1_{- 3.2}^{+3.4}$ \\
& $T^{\rm sd}_{\rm eff}$ & $35841_{-131}^{+131}$ & $35839^{+127}_{-125}$&$35853_{-139}^{+139}$ & $35845_{-139}^{+143}$ \\
& $\log g_{\rm sd}$ [cm s$^{-2}$] & $5.328_{-0.012}^{+0.014}$ &$5.312_{- 0.020}^{+ 0.018}$ & $5.347_{-0.016}^{+0.018}$&$5.341_{-0.020}^{+0.019}$ \\
& $R_{\rm sd}$ ($R_\odot$) & $0.272\pm{0.007}$ &$0.275\pm{0.007}$ &$0.274\pm{0.007}$&$0.275\pm{0.007}$ \\
\hline
\multirow{8}{*}{Derived}
&$q$& $2.18_{-0.17}^{+0.16}$ & $2.24_{-0.21}^{+0.24}$& $1.98_{-0.18}^{+0.18}$&$1.92_{-0.19}^{+0.22}$ \\
& $M_{\rm sd}$ ($M_\odot$) & $0.577_{-0.037}^{+0.039}$  &$0.563_{-0.039}^{+0.039}$ &$0.609_{-0.040}^{+0.042}$ &$0.603_{-0.040}^{+0.042}$ \\
&$M\comp$ ($M_\odot$) & $1.258_{-0.068}^{+0.066}$ & $1.258_{-0.083}^{+0.095}$& $1.203_{-0.077}^{+0.078}$ & $1.158_{-0.083}^{+0.091}$ \\
&$M_{\rm all}$ ($M_\odot$) & $1.834_{-0.083}^{+0.083}$ & $1.823_{-0.092}^{+0.099}$ & $1.812_{-0.088}^{+0.089}$
& $1.765_{-0.092}^{+0.095}$\\
&$a$ ($R\odot$) & $1.468_{-0.022}^{+0.022}$ & $1.465_{-0.025}^{+0.026}$
& $1.462_{-0.024}^{+0.024}$&
$1.449_{-0.026}^{+0.026}$\\
&$L_{\rm sd}$ ($L_\odot$) & $110.2_{-5.7}^{+5.8}$ & $112.0_{-5.8}^{+6.1}$ &$111.5_{-5.9}^{+6.0}$ & 
$112.4_{-6.1}^{+6.0}$\\
&$R^{\rm Roche-lobe}_{\rm sd}$ ($R_{\odot}$) & $0.461_{-0.011}^{+0.011}$ & $0.457_{-0.012}^{+0.011}$ & $0.470_{-0.011}^{+0.012}$ & 
$0.470_{-0.012}^{+0.012}$\\
&$R^{\rm Roche-lobe}\comp$ ($R_{\odot}$) & $0.658_{-0.015}^{+0.014}$ &  $0.659_{-0.019}^{+0.022}$&$0.642_{-0.018}^{+0.018}$ & 
$0.632_{-0.020}^{+0.022}$\\
\hline
\hline
\end{tabular}
\label{tab:diff_prior}
\end{table*}



\begin{table*}
\centering
\small{\textbf{Supplementary Table 4: Natures of compact companion of Lan\,11 as revealed  by our binary population synthesis simulations.}}
\begin{center}
   \begin{tabular}{cccccccccccc}
\hline \hline
\hline
$\alpha_{\rm CE}$ & $\rm CO\,WD$ & $\rm ONe\,WD$ & $\rm NS$\\
\hline
$0.5$ & $7\%$ & $90\%$ & $3\%$\\
$0.75$ & $11\%$ & $87\%$ & $2\%$\\
$1.0$ & $9\%$ & $85\%$ & $6\%$\\
\hline 

\hline \label{1}
\end{tabular}
\end{center}
\end{table*}




\begin{figure*}
    \begin{center}
    \includegraphics[scale=1.15,angle=0]{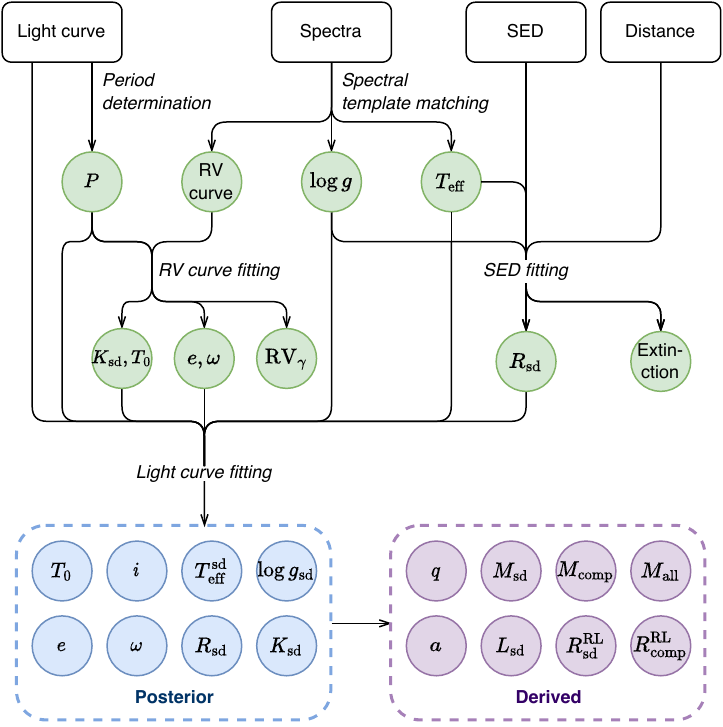}
    \end{center}
    \caption{\textbf {Flowchart of estimates of stellar properties and orbital parameters for Lan\,11.}}
    \label{fig:flowchart}
\end{figure*}

\begin{figure*}
\begin{center}
    \includegraphics[width=2\columnwidth]{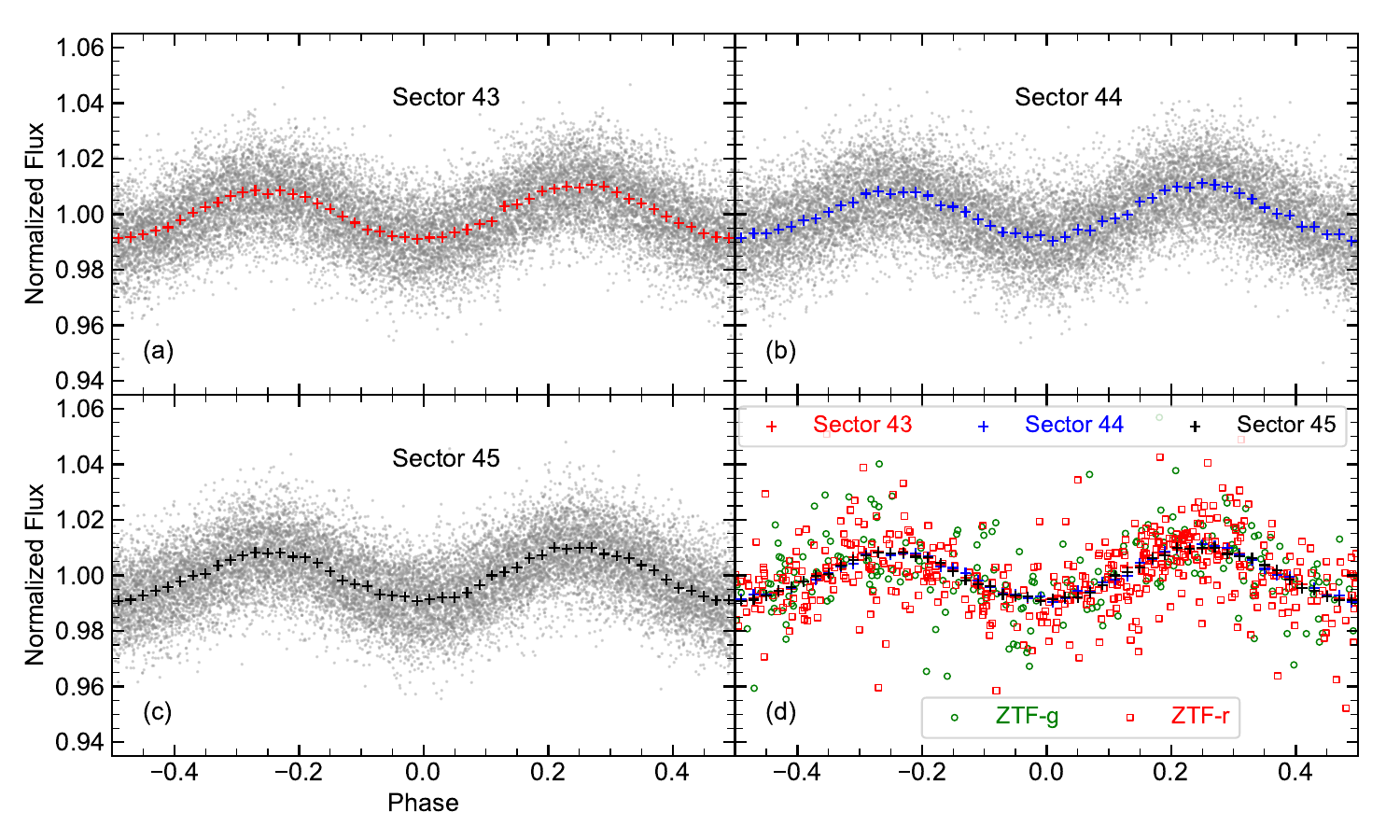}
\end{center}
    \caption{{ \textbf{Phase folded TESS and ZTF light curves.} Panels (a)-(c) show the  phase folded light curves for TESS Sector 43, 44 and 45, respectively. 
    The cross symbols represent the median values of the TESS fluxes within bins from phase $-0.5$ to 0.5 at a step 0.02.  
    Panel (d) shows the light curves for ZTF $g$ (green circle) and $r$ (red square), respectively. The median TESS light curves are also overplotted for comparisons.}}
    \label{fig:lc_tess_sigle_sector}
\end{figure*}


\begin{figure*}
    \begin{center}
    \includegraphics[width=\columnwidth]{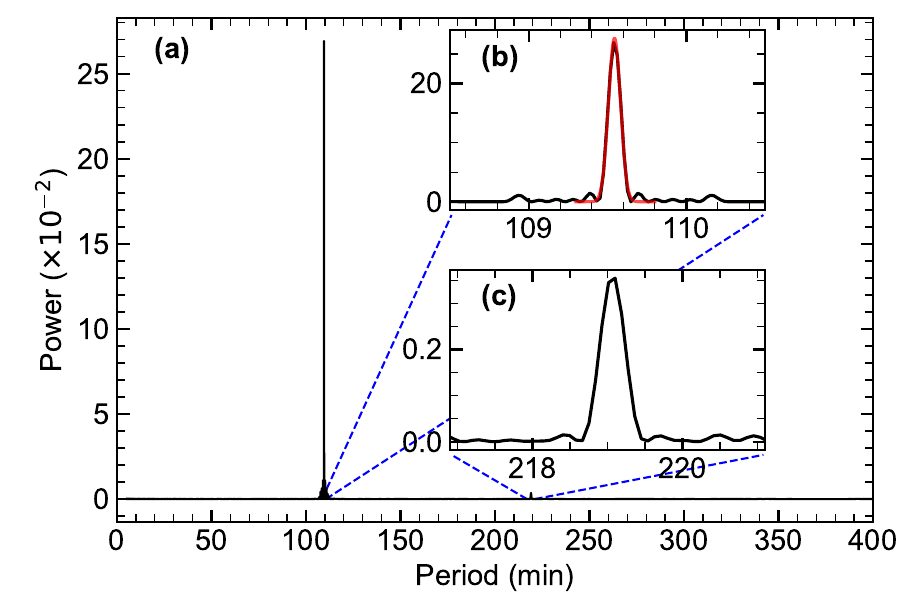}
    \end{center}
    \caption{\textbf{The Lomb–Scargle periodogram of the TESS light curve.} In the full periodogram, a dominant peak can be seen at 109.54440(20) min, corresponding to half of the orbital period. {\it Panel (b)}: the red line is the best-fit Gaussian function using non-linear least squares to fit the periodogram data around the peak period. The error is the square root of the estimated covariance. { {\it Panel (c)}: the peak of the second harmonic arising from the Doppler beaming effect.}}
    \label{fig:lc_period}
\end{figure*}

\begin{figure*}
    \begin{center}
    \includegraphics[scale=0.2,angle=0]{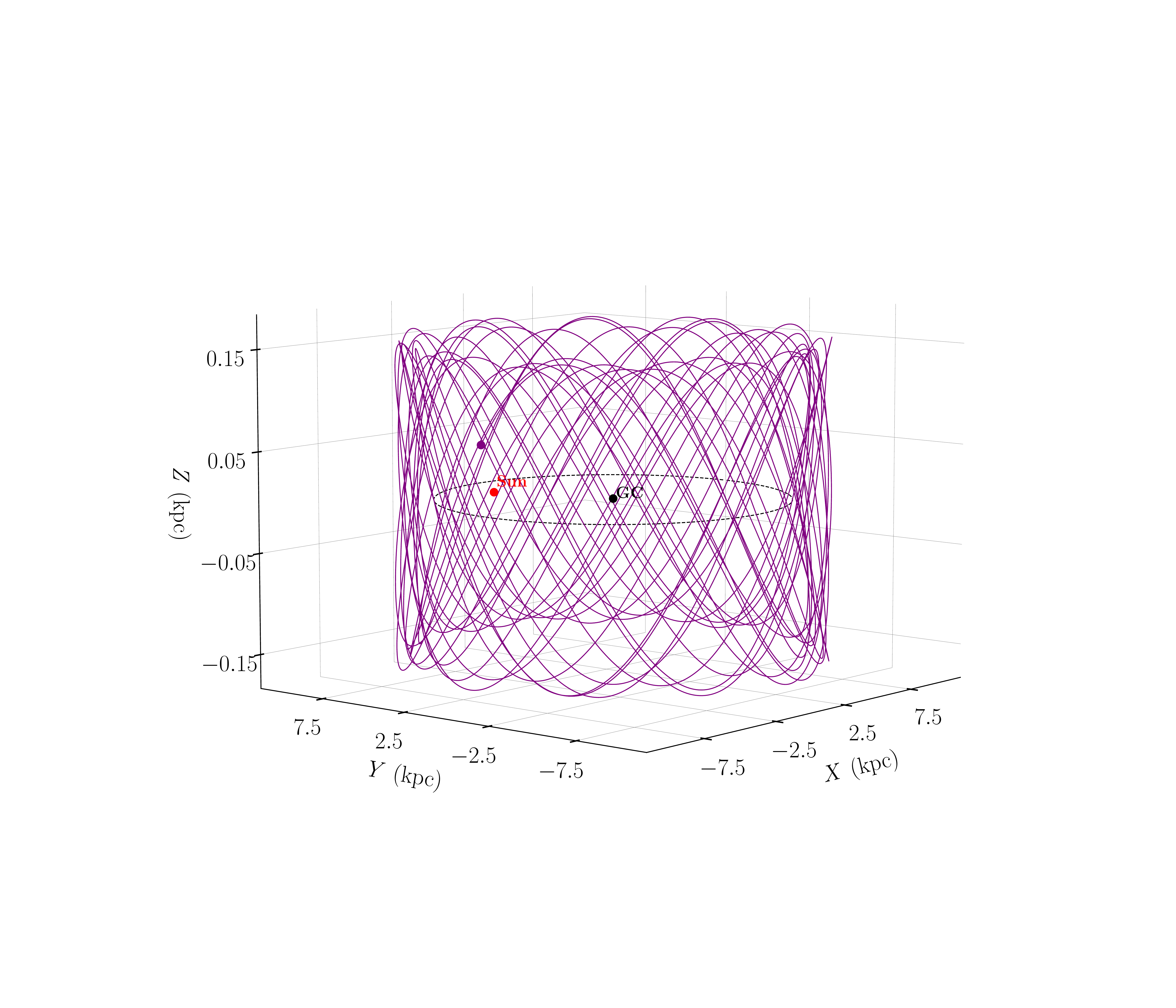}
    \end{center}
    \caption{{ \textbf{3D representation of the backward orbit of Lan\,11.} The orbit is integrated up to 5\,Gyr back in time with a step of 0.1\,Myr. The purple dot marks the current positions of Lan\,11. The red and black dots represent the positions of the Sun and the Galactic Center, respectively. 
    The Solar circle ($R = 8.178$\,kpc) is marked by the dotted gray lines.}}
    \label{fig:3d_orbit}
\end{figure*}

\begin{figure*}
    \begin{center}
    \includegraphics[scale=0.55, angle=0]{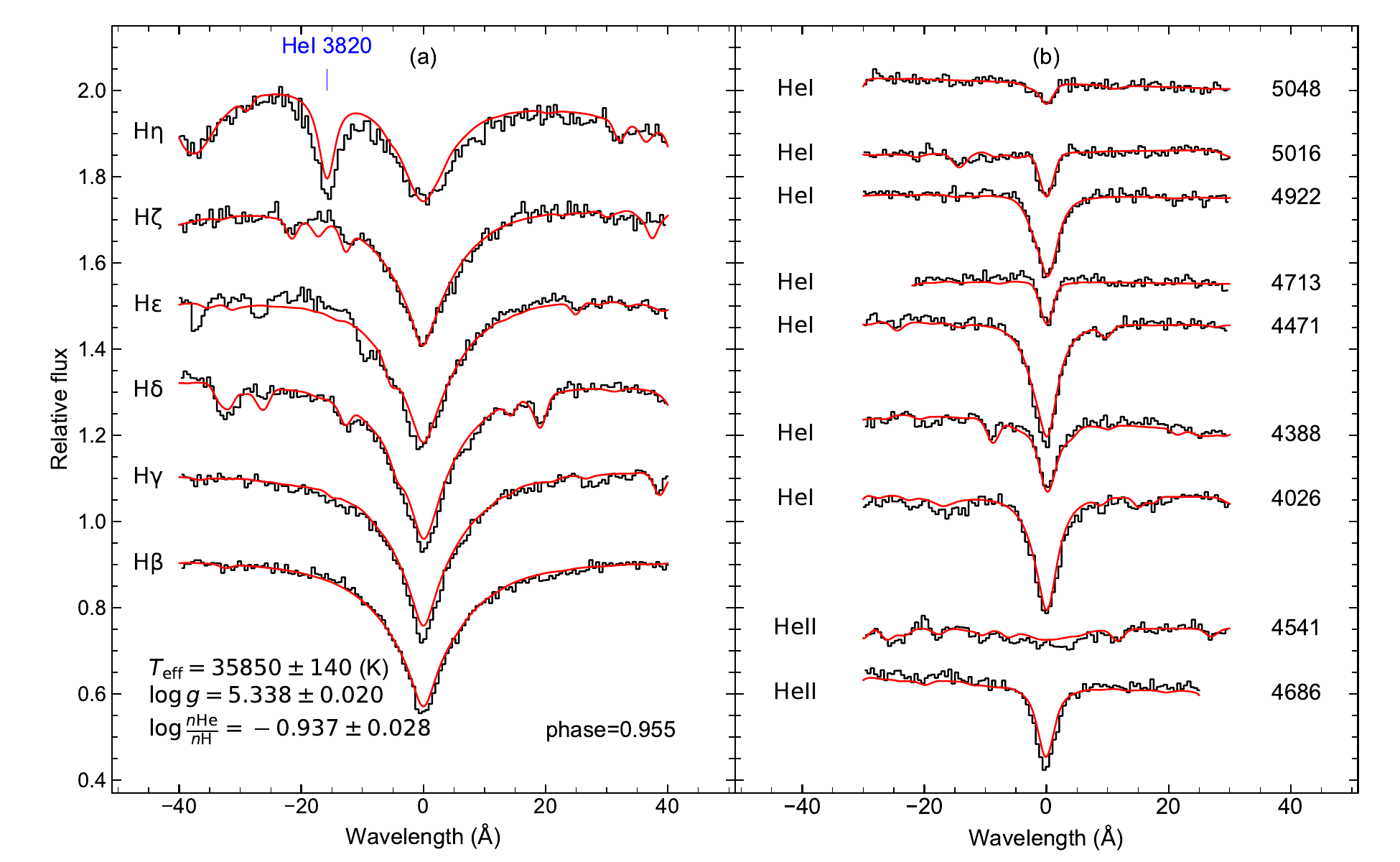}
     \end{center}
    \caption{\textbf {Spectral fit to highest SNR DBSP spectrum.} Panels (a) and (b) show the fit to the hydrogen and helium lines of  {the highest SNR DBSP spectrum} observed at the orbital phase of 0.955.
    The observed spectral lines are shown in black, and the best-fit model lines are shown in red.  {The best-fit parameters are shown in the bottom-left corner of panel (a).}}
    \label{fig:dbsp_lines}
\end{figure*}

\begin{figure*}
    \begin{center}
    \includegraphics[scale=0.55, angle=0]{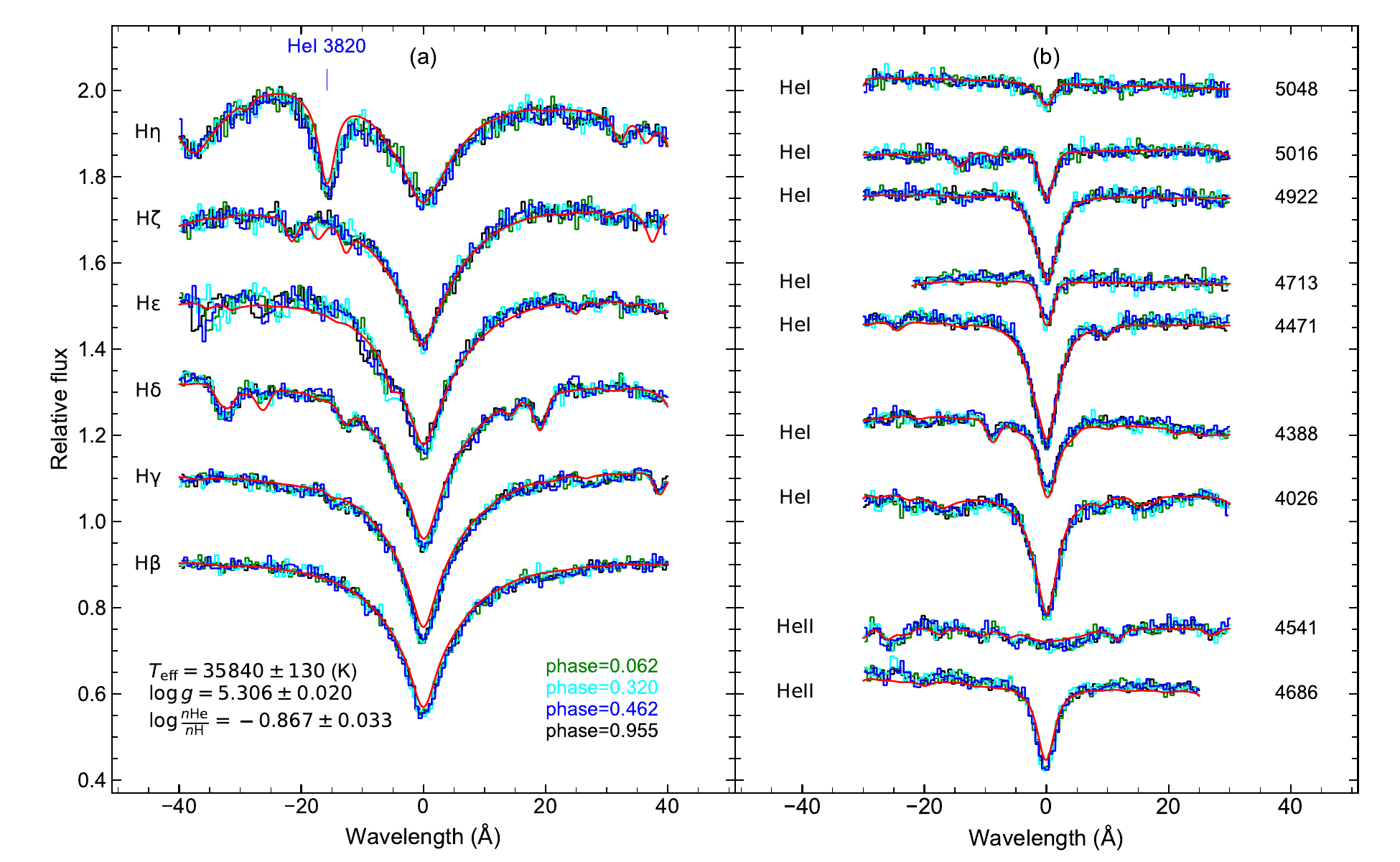}
     \end{center}
    \caption{\textbf {Spectral fit to four phase-picked DBSP spectra.} Panels (a) and (b) show the fit to the hydrogen and helium lines of  {four phase-picked high-SNR spectra. The orbital phase of each spectrum is marked in the bottom-right corner of panel (a)}.
    The observed spectral lines are shown in green, cyan, blue, and black, and the best-fit model lines are shown in red.  {The best-fit parameters are shown in the bottom-left corner of panel (a).}}
    \label{fig:dbsp_lines_phase_picked}
\end{figure*}


\begin{figure*}
\begin{center}
    \includegraphics[width=1.7\columnwidth]{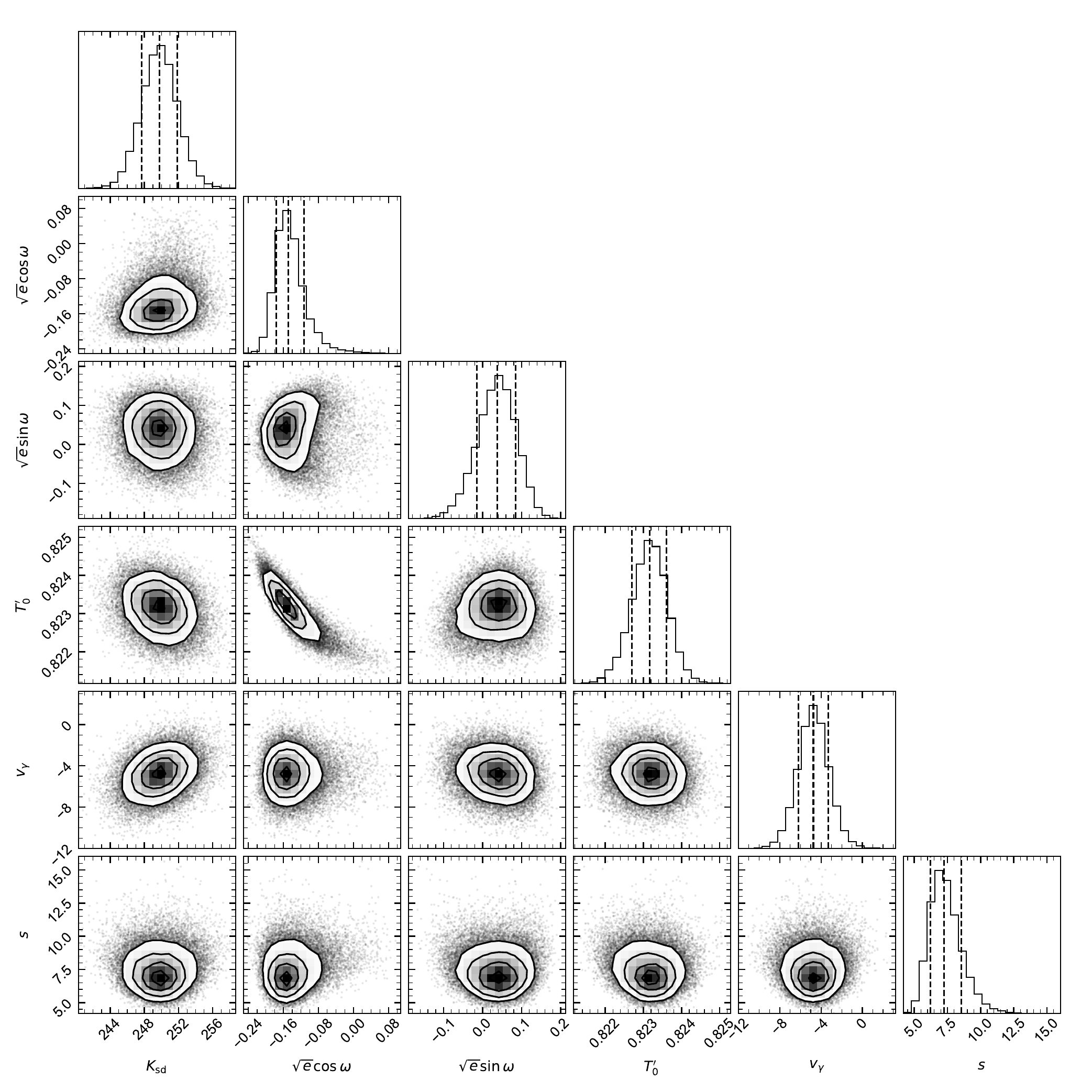}
\end{center}
    \caption{ \textbf{Corner plot for the MCMC sampling of the radial velocity curve.} Posterior probability distribution of the parameters directly derived by the MCMC simulation based on the radial velocity curve measured from the DBSP spectra. Note that $T^\prime_{0} = T_{0} - 2459477$ is in BJD, $K_{\rm sd}$, $v_{\gamma}$ and $s$ are in \kms, and $s$ is the jitter. }
    \label{fig:rv_coner}
\end{figure*}

\begin{figure*}
    \includegraphics[width=1.6\columnwidth]{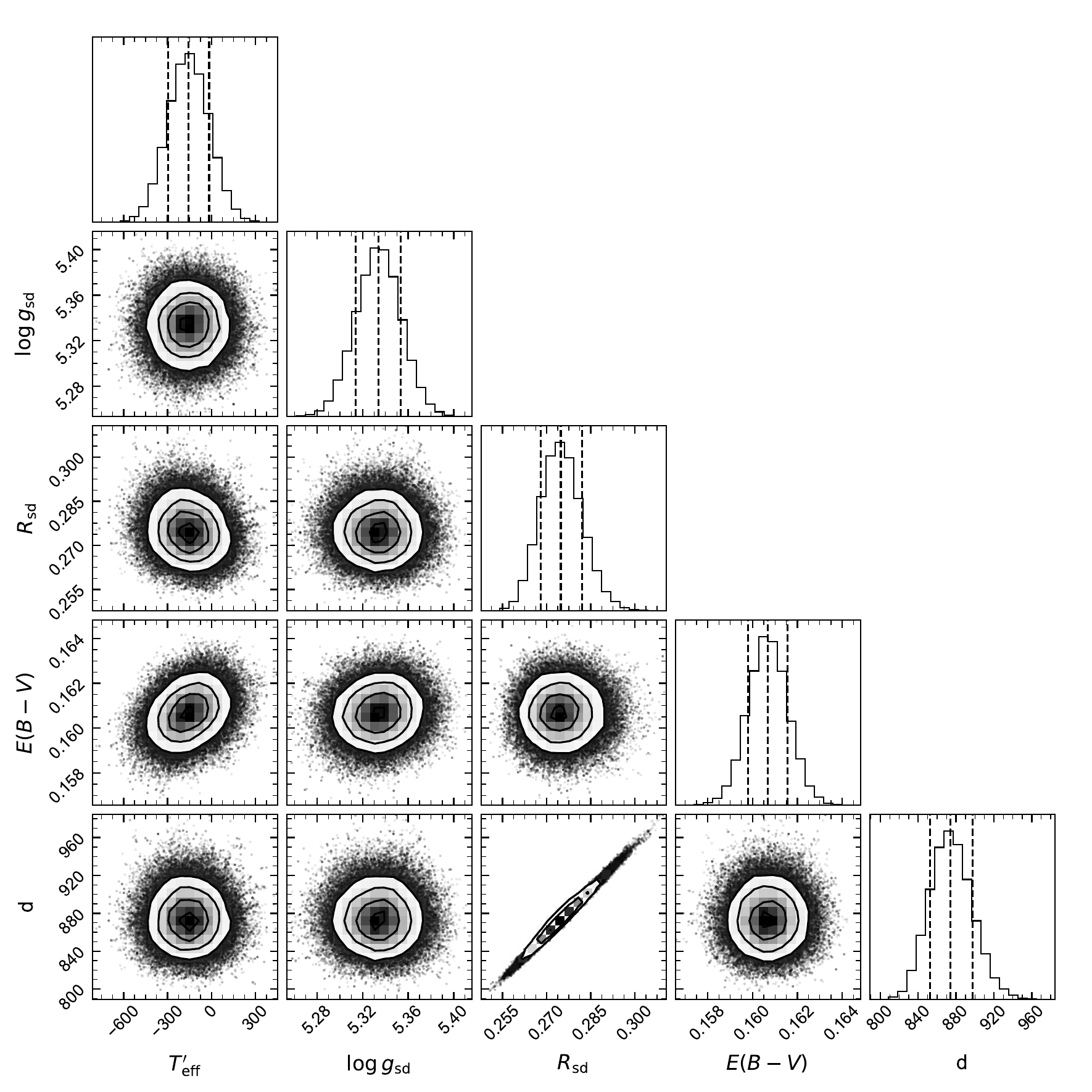}
    \caption{{\textbf{Corner plot for the SED MCMC sampling using {\tt speedyfit}.} Posterior probability distribution of the parameters directly derived by the MCMC simulation based on the multi-band photometry, including {\it Gaia} EDR3 $G_{\rm bp}$-, $G$- and $G_{\rm rp}$-bands, PanSTARRS $z$- and $y$-bands, 2MASS $J$-, $H$- and $K_{\rm s}$-bands, and ALLWISE $W_1$-, $W_2$-bands. The effective temperature of hot subdwarf $T_{\rm eff} = T^\prime_{\rm eff} + 36000$\,K. 
    $R_{\rm sd}$ is the radius of the hot subdwarf in solar radius; log$g_{\rm{ sd}}$ is the surface gravity of hot subdwarf star, $E(B-V)$ is extinction in magnitude, and $d$ is the distance in parsec.}}
    \label{fig:SED_corner}
\end{figure*}

\begin{figure*}
\begin{center}
    \includegraphics[width=2.1\columnwidth]{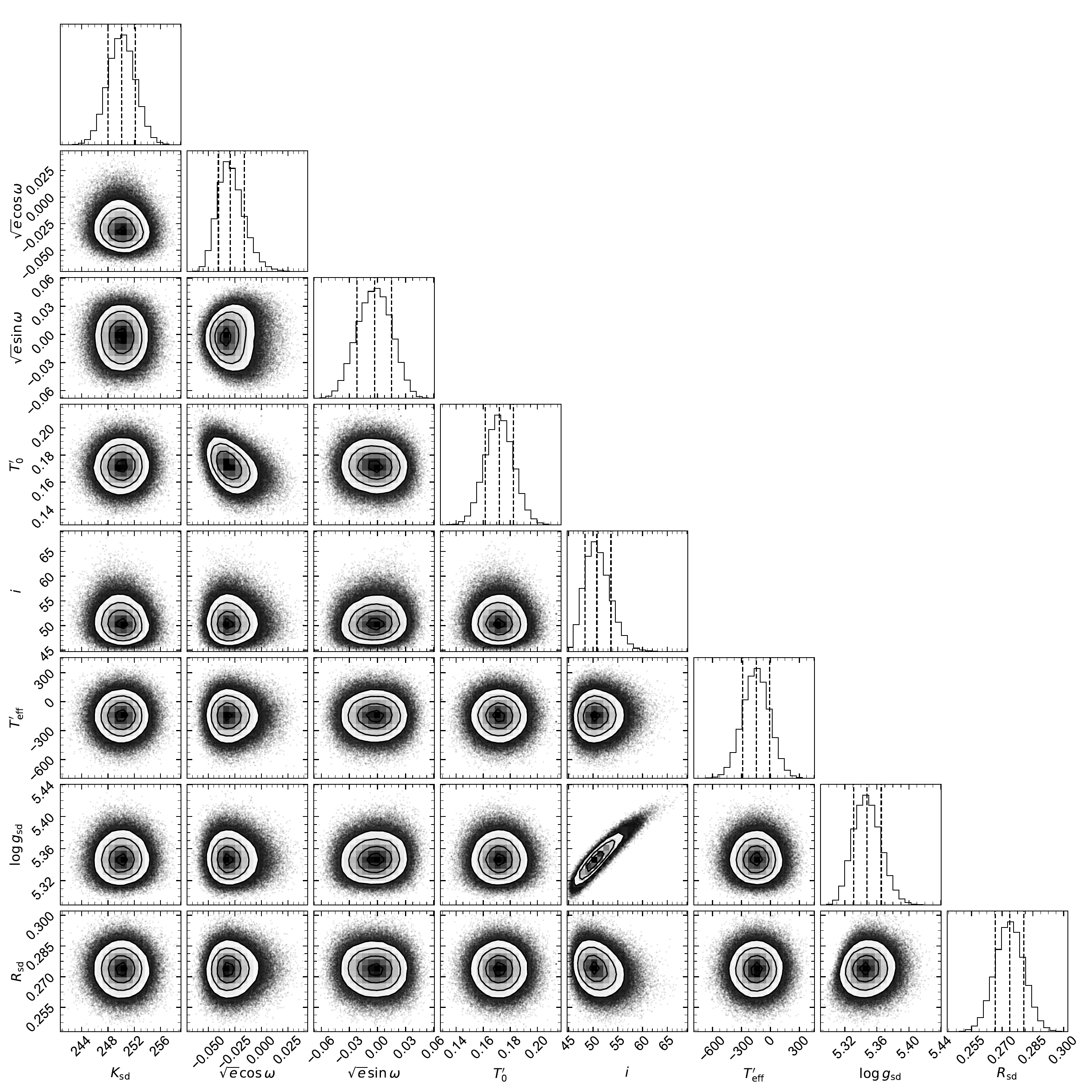}
\end{center}
    \caption{{\textbf{Corner plot for the MCMC sampling of the light curve using \texttt{ellc}.} Posterior probability distribution of the parameters directly derived by the MCMC simulation based on the \texttt{TESS} light curve. Labels shown are the semi-amplitude radial velocity of the hot subdwarf $K_{\rm sd}$ in \kms, the longitude of periastron $\omega$ in degree, eccentricity $e$, the superior conjunction time $T_{\rm 0}$ in BJD, $T^\prime_{\rm 0} = (T_{\rm 0} - 2459477.82) \times 100$, binary orbital inclination angle $i$ in degree, the effective surface temperature of the hot subdwarf $T_{\rm eff} = T^\prime_{\rm eff} + 36000$\,K, the surface gravity log$g_{\rm{ sd}}$, and the radius of hot subdwarf $R_{\rm sd}$ in solar radius. 
    }}
    \label{fig:ellc_coner}
\end{figure*}


\begin{figure*}
\begin{center}
    \includegraphics[width=2\columnwidth]{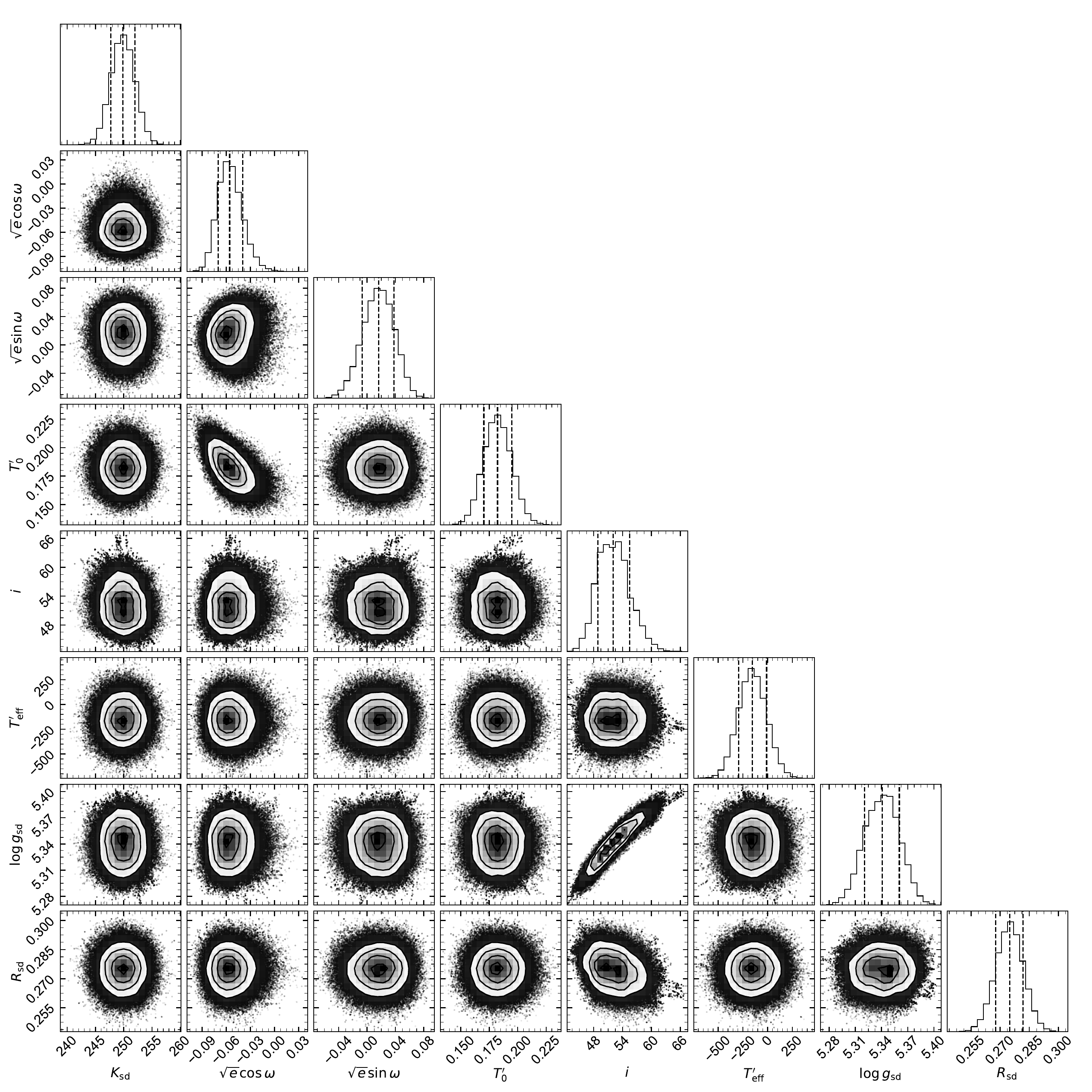}
\end{center}
    \caption{{\textbf{Corner plot for the MCMC sampling of the light curve using {\tt PHOEBE}.} Posterior probability distribution of the parameters directly derived by the MCMC simulation based on the TESS light curve using {\tt PHOEBE}. 
    The labels are same as these in Supplementary Figure\,9.}}
    \label{fig:phoebe_coner}
\end{figure*}

\begin{figure*}
\begin{center}
    \includegraphics[width=\columnwidth]{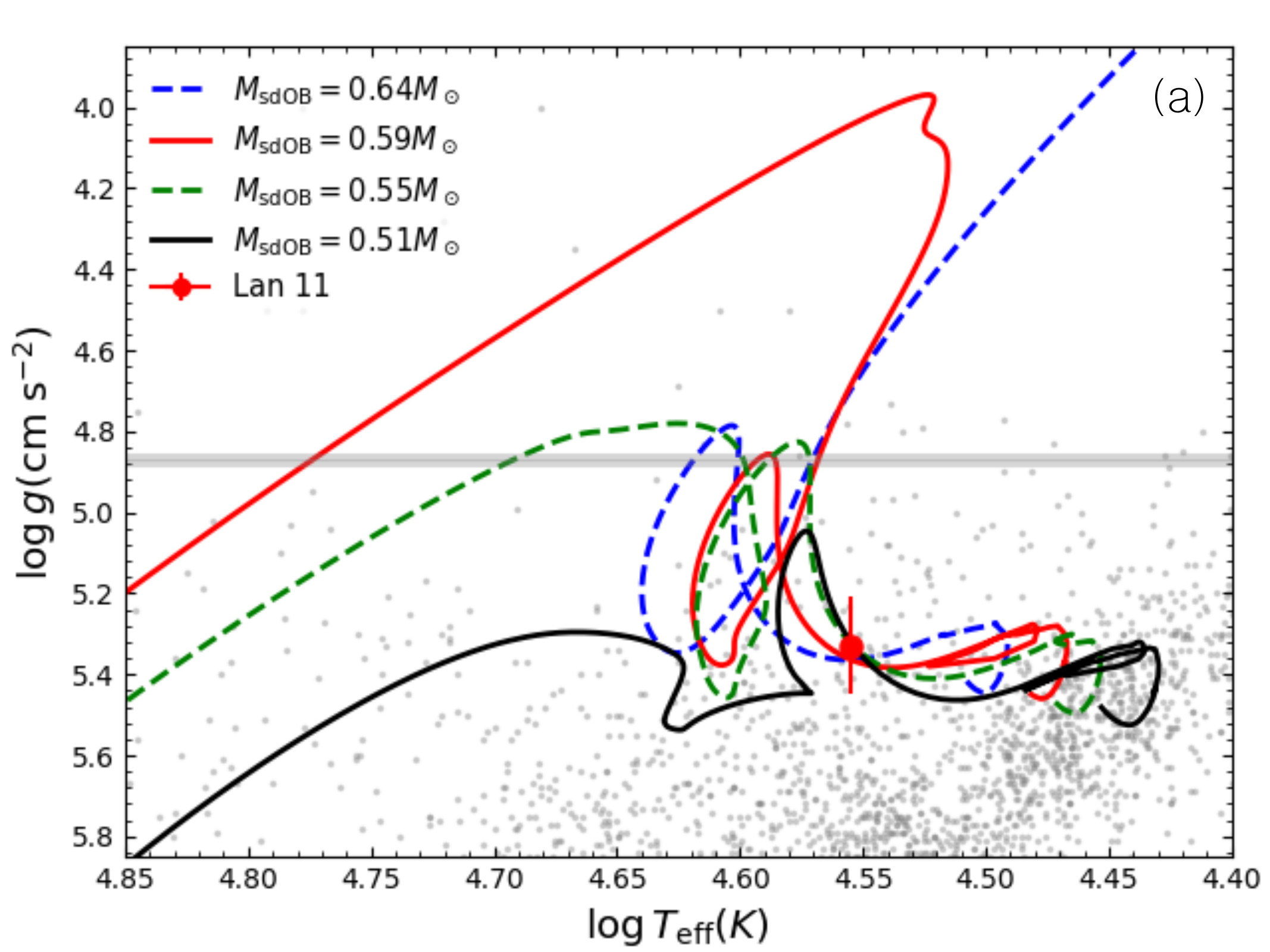}
    \includegraphics[width=\columnwidth]{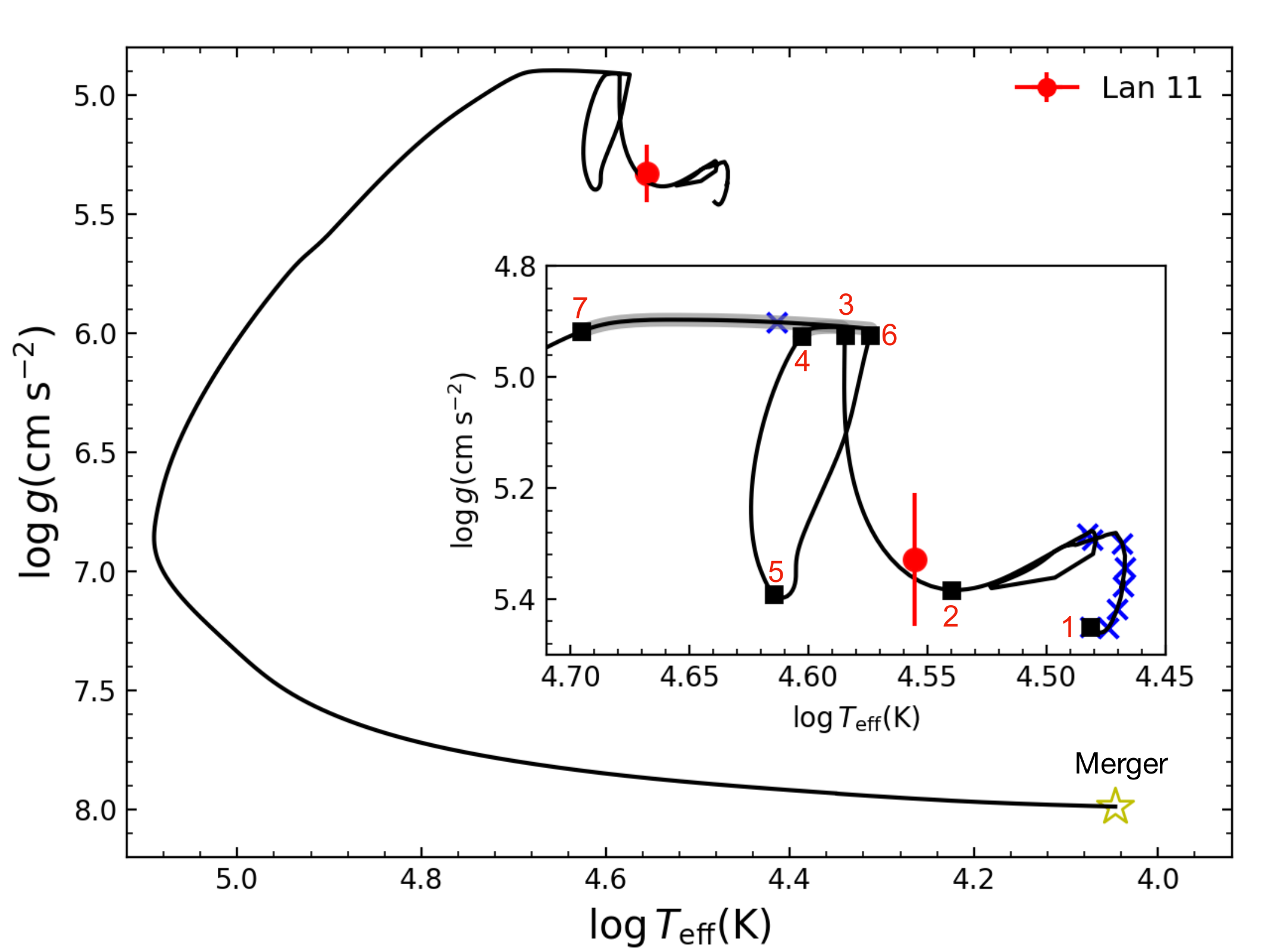}
\end{center}
    \caption{{ \textbf{The evolutionary tracks of a LAN\,11-like sdOB-WD binary system.} Panel a: Hot subdwarf tracks with different masses: $M_{\rm sdB}=$ $0.51$, $0.55$, $0.59$, $0.64$ $M_\odot$ (with corresponding progenitor masses: $2.6$, $2.8$, $3.0$, $3.2$ $M_\odot$).
    The location of Lan\,11 is shown in red dot, and the grey dots represent the observational sample of sdOB stars taken from ref.\citet{Geier2015}. 
    The grey line represents the approximate position of $\log g$ if the sdOB star radius approaches to the Roche lobe radius. 
    It is noted that the sdOB star with mass of $0.51M_\odot$ would never fills its Roche lobe. 
    Panel b: The evolutionary details of a Lan\,11-like system with $M_{\rm sdB}=0.59M_\odot$ and a compact companion of $1.29M_\odot$.} The track starts from He core burning, and ends at the merger event, as shown in the yellow star. In the inset, the blue crosses mark the time interval of $10^7\;\rm yr$. The black squares represent different evolution states: He core burning (1), H burning in the shell (2), the onset of the first mass transfer phase (3), the termination of the first mass transfer phase (4), the He burning in the shell (5), the onset of the second mass transfer phase (6), and the termination of the mass transfer phase (7). The mass transfer phase is shown in a thick black line.}
    \label{fig:track}
\end{figure*}

\section*{Code Availability}
We use standard data analysis tools in the Python environments. Specifically, the Galactic orbital analysis is carried out with Python package \href{https://gala-astro.readthedocs.io/en/latest/#}{Gala}\cite{Price-Whelan17}. TESS light curve is extracted and downloaded with \href{https://docs.lightkurve.org/#}{{\tt lightkurve}}\cite{lightkurve}. Light curve modelling is performed with packages: \href{http://www.phoebe-project.org/releases/2.4}{\tt phoebe} 2.4.7 and \href{https://github.com/pmaxted/ellc}{\tt ellc}\cite{Maxted16}. 
MCMC is performed with package \href{https://emcee.readthedocs.io/en/stable/index.html}{\tt emcee}\cite{emcee2013} 3.1.3. 
The binary orbital radial velocity is calculated by \href{https://radvel.readthedocs.io/en/latest/}{\tt radvel}\cite{Fulton2018}.
{\tt mpfit} is available at \url{https://www.l3harrisgeospatial.com/docs/mpfit.html}.
The Python packages \href{https://www.astropy.org/}{\tt astropy}\cite{astropy2013, astropy2018}, \href{https://numpy.org/}{\tt numpy}\cite{Numpy2011CSE, Numpy2020Natur} and  \href{https://scipy.org/}{\tt scipy}\cite{scipy2001jones, scipyNMeth2020} are also used.
The DBSP spectra are extracted by \href{https://github.com/lidihei/pyexspec}{{\tt pyexspec}}, 
their radial velocity is measured using package \href{https://github.com/hypergravity/laspec}{{\tt Laspec}}\cite{Zhangbo2020}. The SED fitting is performed by using \href{https://speedyfit.readthedocs.io/en/stable/}{\tt SPEEDYFIT}. 
All module mentioned above are publicly available. 
{\sc XTgrid} is publicly available online at \url{www.Astroserver.org}.
The Hurley rapid binary evolution code\cite{HUR00, HUR02} is publicly available online at \url{https://astronomy.swin.edu.au/~jhurley/}.
The binary evolutionary model is carried out with \href{https://docs.mesastar.org/en/release-r22.11.1/}{MESA} 12115\cite{Paxton11}.

\section*{References}
\bibliographystyle{naturemag-doi}
\bibliography{refs.bib}

\newpage

\begin{addendum} 

\item[Acknowledgements] 
We acknowledge the staff of the XLT and Palomar observatories for assistance with the observations. We thank Jia-Lu Nie and Dr. Yu-Jiao Yang for Palomar remote observatories. We thank Prof. Xiao-Bing Zhang for discussions. This work is supported by the National Natural Science Foundation of China (NSFC) for grants No. 11988101, No. 11933004, No. 12288102, No.12125303, No. 12090040, and No. 12225304. This work is also supported by the National Key R\&D Program of China grants No. 2019YFA0405500, No. 2021YFA1600400, and No. 2021YFA1600401, the China Manned Space Project with no. CMS-CSST-2021-A08 and CMS-CSST-2021-A10. Z.W.L. acknowledges support from NSFC grant No. 12103086 and the Yunnan Fundamental Research Projects (YFRP) grant Nos. 202201AU070234 and 202301AT070314. D.D.L. is supported by  NSFC grant No. 12273105, the Youth Innovation Promotion Association CAS grant No. 2021058, the Western Light Project of CAS (No. XBZG-ZDSYS-202117), and the YFRP grants No. 202301AV070039, No. 202101AT070027 and No. 202101AW070047. Y.P.L. acknowledges support from NSFC grant no. 12173028. B.Z. acknowledges the suport from NSFC under grant No.12203068. P.W. acknowledges support from the NSFC under grant No. U2031117, the Youth Innovation Promotion Association CAS (id. 2021055), CAS Project for Young Scientists in Basic Reasearch grant No. YSBR-006 and the Cultivation Project for FAST Scientific Payoff and Research Achievement of CAMS-CAS. P.N. acknowledges support from the Grant Agency of
the Czech Republic (GA\v{C}R 22-34467S).
The Astronomical Institute in Ond\v{r}ejov is supported by the project RVO:67985815.
X.F.C. also acknowledges support from the NSFC for grants No. 12288102, No. 12125303. and No. 12090043, and the International Centre of Supernovae, Yunnan Key Laboratory (No. 202302AN360001), the Yunnan Fundamental Research Projects (grant Nos. 202201BC070003, 202001AW070007) and the ``Yunnan Revitalization Talent Support Program'' -- Science \& Technology Champion Project (No. 202305AB350003).
J.F.L. acknowledges support the NSFC through grant Nos. of 11988101 and 11933004, and support from the New Cornerstone Science Foundation
through the New Cornerstone Investigator Program and the XPLORER PRIZE.

Guoshoujing Telescope (the Large Sky Area Multi-Object Fiber Spectroscopic Telescope, LAMOST) is a National Major Scientific Project built by the Chinese Academy of Sciences. The National Development and Reform Commission has provided funding for the project. LAMOST is operated and managed by the National Astronomical Observatories, Chinese Academy of Sciences.

We used data from the European Space Agency mission Gaia (http://www.cosmos.esa.int/gaia), processed by the Gaia Data Processing and Analysis Consortium (DPAC; see \url{http://www.cosmos.esa.int/web/gaia/dpac/consortium}). We also used the data from the SDSS survey.

This paper includes data collected by the TESS mission. Funding for the TESS mission is provided by the NASA Explorer Program. This work presents results from the European Space Agency (ESA) space mission Gaia. Gaia data are being processed by the Gaia Data Processing and Analysis Consortium (DPAC). Funding for the DPAC is provided by national institutions, in particular the institutions participating in the Gaia Multi Lateral Agreement (MLA).


This research has used the services of \mbox{\url{www.Astroserver.org}} under reference OL4CLY. 

\item[Author Contributions]
C.Q.L. discovered this system and led the follow-up observations, wrote parts of the manuscript;
J.L. led the analysis of the light curve and spectra, wrote parts of the manuscript;
C.J.Z. led the analysis of the light curve modelling and SED fitting, wrote parts of the manuscript;
D.D.L led the binary population synthesis simulation;
Z.W.L. performed the MESA numerical simulations to investigate the life of this system;
Y.P.L. assisted with the determinations of stellar parameters of the hot subdwarf;
P.N. determined the stellar parameters of the hot subdwarf using the XT{\tt GRID} tool;
B.Z. assisted with the data reduction of P200;
J.P.X. helped the remote observations;
B.W. joined the discussions and contributed to the revisions of the text;
S.W. helped the remote observations;
Y.B. helped the remote observations;
Q.Z.L. performed the kinematic analysis of this system;
P.W. performed the data reduction and analysis of FAST observations;
Z.W.H. helped the interpretation of the results;
J.F.L. contributed to the interpretation of the results and revised the text;
Y.H. wrote the manuscript and partly led the project;
X.F.C. led the interpretation of the results and revised the text;
C.L. led the project, organized the observations and revised the text.

\item[Competing interests statement]
The authors declare no competing interests.

\item[Author Information] 
Correspondence and requests for materials should be addressed to Yang Huang, Xuefei Chen and Chao Liu.

\end{addendum}

\end{document}